\documentclass[11pt]{article}
\usepackage{epsfig,amssymb,amsmath,amsthm,graphicx,mathtools,
xcolor,multirow,booktabs,hyperref}

\usepackage[sectionbib,longnamesfirst]{natbib}
\setcitestyle{authoryear}
\advance\topmargin by -60pt
\newcommand{\mcD}{\mathcal{D}}
\newcommand{\bfbeta}{\mbox{\boldmath $\beta$}}
\newcommand{\bftheta}{\mbox{\boldmath $\theta$}}
\newcommand{\bfvartheta}{\boldsymbol{\vartheta}}
\newcommand{\bfeta}{\boldsymbol{\eta}}
\newcommand{\bfdelta}{\boldsymbol{\delta}}

\numberwithin{equation}{section}

\begin{document}

\begin{center}
\large{\textbf{ON INFORMATION ABOUT COVARIANCE PARAMETERS IN GAUSSIAN MAT\'ERN RANDOM FIELDS}}

\vspace{1cm}

Victor De Oliveira \\
Department of Management Science and Statistics \\
The University of Texas at San Antonio, U.S.A. \\
{\tt victor.deoliveira@utsa.edu}

\vspace{0.5cm}

Zifei Han\footnote{Corresponding author.} \\
School of Statistics \\
University of International Business and Economics, China \\
{\tt zifeihan@uibe.edu.cn}

\vspace{0.3in}

 Aug 29, 2022 \\
(final version)
\end{center}

\vspace{0.1in}

{\small
\begin{center}
\textbf{Abstract}
\end{center}

\noindent
The Mat\'ern family of covariance functions is currently the most commonly used for the 
analysis of geostatistical data due to its ability to describe different smoothness behaviors. 
Yet, in many applications the smoothness parameter is set at an arbitrary value. 
This practice is due  partly to computational challenges faced when attempting to
estimate all covariance parameters and partly to unqualified claims in the literature stating that 
geostatistical data have little or no information about the smoothness parameter.
This work critically investigates this claim and shows it is not true in general. 
Specifically, it is shown that the information the data have about the correlation parameters
varies substantially depending on the true model and sampling design and, in particular,
the information about the smoothness parameter can be large, in some cases larger than 
the information about the range parameter. In light of these findings, we suggest to reassess the aforementioned practice and instead establish inferences from 
data--based estimates of both range and smoothness parameters, 
especially for strongly dependent non--smooth processes observed on irregular sampling designs. 
A data set of daily rainfall totals is used to motivate the discussion and gauge this common practice.
\vspace{0.2in}

}

\vspace{0.2in}

\newpage

\section{Introduction}  \label{sec:intro}

Random fields are ubiquitous for the modeling of spatial and spatio--temporal data
in most natural and earth sciences, such as ecology, epidemiology,  geology and hydrology.
Among these, Gaussian random fields play a prominent role due to their versatility to model
spatially varying phenomena, and because they serve as building blocks for the construction 
of more sophisticated models \citep{Zimmerman2010,Gelfand2016}.
One of the common scientific goals in these sciences is spatial 
interpolation/prediction, and for this the covariance function of the random field plays a key role. 
Reliable modeling and inference of covariance functions of Gaussian random fields 
are then crucial steps toward this goal.

A large number of parametric families of covariance functions have appeared in the 
statistical literature, but just a few of them are commonly used in practice. 
One family that has reached prominence is the so--called {\it Mat\'ern} family \citep{Matern1986}.
Except for a few sporadic appearances, this family was introduced in the statistical literature 
by \cite{Handcock1993}, and later \cite{Stein1999} studied its properties and strongly advocated 
its use; see \cite{Guttorp2006} for the history of this family.
There are two main reasons that explain this prominence.
First, unlike most other families that are indexed by a variance and a range parameter, 
covariance functions in the Mat\'ern family also depend on an additional parameter, 
called the smoothness parameter, that controls the degree of mean square differentiability of 
the random field. 
Second, in a series of articles \cite{Stein1988,Stein1990,Stein1993}
established that,  
in the so--called fixed--domain asymptotic framework, it is possible to achieve efficient spatial 
interpolation with a misspecified covariance model, as long as the correct and misspecified models
are compatible on the region of interest in some well defined sense.
A necessary condition for two covariance functions from the Mat\'ern family 
to be compatible in that sense is that they share the same smoothness parameter.
The other parameters may differ, as long as these satisfy a certain relation \citep{Zhang2004}. 
This suggests that for efficient spatial interpolation/prediction using the Mat\'ern family, 
correct specification of the smoothness parameter is critical and may even be more important
than correct specification of the other parameters, at least in this particular asymptotic framework.
The same situation has been recently found to hold for other families of covariance functions 
\citep{Bevilacqua2019}. 

In spite of the above, in some applied works there is a disconnect between
theory and practice when it comes to modeling geostatistical data using the Mat\'ern family.
On the one hand, the facts stated above establish the importance of an adequate specification of the smoothness of the random field when the main goal is spatial prediction, which has been advocated by
several researchers \citep{Stein1999}. 
On the other hand, the smoothness of the random field is 
oftentimes arbitrarily fixed in advance in these applied works rather than estimated. 
Even though there is usually little or no a priori information about the smoothness of the random field 
of interest, the exponential covariance family is commonly used as the default model
(a sub--family of the Mat\'ern family) which is not mean square differentiable. 
This imposes from the onset a lack of smoothness in the random field, which may or may not be 
supported by the data.
But gross misspecification of the smoothness of the random field likely precludes 
the possibility of efficient spatial interpolation/prediction. 
The same situation has been recently found to hold for other covariance families \citep{Bevilacqua2019}.
This practice, somewhat common in geostatistics when estimating covariance functions,
is motivated partly by numerical challenges practitioners sometimes face when
attempting to estimate all parameters in the Mat\'ern family, and partly by an unqualified claim in the literature stating that geostatistical data have little or no information about the smoothness of the random field.
The latter has become part of the `geostatistical folklore' in some applied quarters.
The main aim of this work is to critically investigate this claim, and contribute to the bridging of theory and practice on this issue.

An area of application where this issue is especially relevant is the spatial modeling of 
rainfall fields.
There are three (main) types of rainfall, depending on the atmospheric mechanism that triggers it:
stratiform (predominant in northern--latitudes), convective (predominant in the tropics) and 
orographic (mountains).
Rainfall from the former type tends to occur over large spatial scales, being of long duration,
and varies slowly over space.
On the other hand, rainfall from the latter two types tends to be more localized, being of 
short duration, with a very high gradient over short distances, with sharp transitions between 
the dry and wet sub--regions \citep{Steiner1995}. 
It is then expected that random fields that describe the spatial variation of stratiform rainfall
to be smoother than random fields describing the spatial variation of convective and orographic rainfall.
Consequently, rather than fixing the smoothness of the random field for rain data, 
a more satisfactory practice would be to estimate it, tailoring it to the 
type of atmospheric mechanism that generated the data. 

Using the inverse diagonal elements of the inverse of the Fisher information matrix 
as a tool, this work investigates the information content that geostatistical data have about the covariance 
parameters of the Mat\'ern family, with emphasis on the correlation parameters and
a particular microergodic parameter. 
It is shown that the information about these parameters can vary widely depending on 
the true model and the sampling design, and the information about the smoothness parameter can be 
substantial, when the information about the range parameter is considered as reference. 
These findings cast doubt on the aforementioned claim, and invite to reassess the practice of
fixing the smoothness parameter at an arbitrary value.
A data set of daily rainfall totals collected in Switzerland  is used to motivate the discussion and 
assess the aforementioned practice.

\section{Data and Model}  \label{subsec:data-models}

Geostatistical data consist of triplets 
$\{({\bf s}_i, \boldsymbol{f}({\bf s}_i), z_i) : i=1,\ldots,n\}$, 
where $\mathcal{S}_n = \{{\bf s}_1,\ldots,{\bf s}_n\}$ is a set of sampling locations in 
the region of interest, called the {\it sampling design}, 
$\boldsymbol{f}({\bf s}_i) = (f_1({\bf s}_i),\ldots, f_p({\bf s}_i))^{\top}$ is a
$p$--dimensional vector of covariates measured at ${\bf s}_i$ 
(usually $f_{1}({\bf s}) \equiv 1$), 
and $z_i$ is a measurement of the quantity of interest collected at ${\bf s}_i$.
The stochastic approach to spatial interpolation/prediction relies on viewing the set of 
measurements $\{z_i\}_{i=1}^{n}$ as a partial realization of a random field.

Let $\{Z({\bf s}) : {\bf s} \in \mcD\}$ be a Gaussian random field with mean function 
$\mu({\bf s})$ and covariance function $C({\bf s},{\bf u})$, 
with $\mcD \subset {\mathbb R}^d$ and $d \geq 1$. 
It would be assumed that $\mu({\bf s}) =  \sum_{j=1}^{p} \beta_{j} f_{j}({\bf s})$, where
$\bfbeta = (\beta_{1},\ldots,\beta_{p})^{\top} \in {\mathbb R}^{p}$ are 
unknown regression parameters.
Additionally, $C({\bf s},{\bf u})$ is assumed isotropic and belonging to a parametric family, 
$\big\{C_{\boldsymbol{\theta}}({\bf s},{\bf u}) = 
\sigma^2 K_{\boldsymbol{\vartheta}}(||{\bf s} - {\bf u}||) 
: \bftheta = (\sigma^2, \mbox{\boldmath $\vartheta$}) \in (0,\infty) \times \Theta \big\}$, 
$\Theta \subset {\mathbb R}^q$, where $K_{\boldsymbol{\vartheta}}(\cdot)$ is an isotropic 
correlation function in ${\mathbb R}^d$ and $\| \cdot \|$ is the Euclidean norm. 
Among the many possible isotropic covariance families, we focus in this work on 
the Mat\'ern family with the parametrization proposed in 
\cite{Handcock1993}

\begin{align} 
\label{eq:matern-cov}
C_{\boldsymbol{\theta}}(r) & =  \frac{\sigma^2}{2^{\nu - 1} \Gamma(\nu)}
\left( \frac{2\sqrt{\nu}}{\vartheta} r \right)^{\nu} 
\mathcal{K}_{\nu}\left( \frac{2\sqrt{\nu}}{\vartheta} r \right)  , \quad\quad r \geq 0  \\
 & =:  \sigma^2 K_{\boldsymbol{\vartheta}}(r), \nonumber
\end{align}
where $r \; = ||\bf s - \bf u||$ is Euclidean distance 
 between two locations, $\sigma^2 > 0$, $\mbox{\boldmath $\vartheta$} = (\vartheta,\nu) \in (0,\infty)^2$ are 
correlation parameters, $\Gamma(\cdot)$ is the gamma function and $\mathcal{K}_{\nu}(\cdot)$ 
is the modified Bessel function of second kind and order $\nu$
(\citealp[8.40]{Gradshteyn2000}).
For this family, $\sigma^2 = {\rm var}(Z({\bf s}))$, 
$\vartheta$ (with units of distance) mostly controls how fast 
$C_{\boldsymbol{\theta}}(r)$ decays to zero when $r$ increases, and 
$\nu$ (unitless) controls the degree of differentiability of 
$C_{\boldsymbol{\theta}}(r)$ at the origin. 
When $\nu > k$, $C_{\boldsymbol{\theta}}(\cdot)$ is $2k$ times differentiable at $r=0$,
which in turn implies that $Z(\cdot)$ is $k$--times mean square differentiable \citep{Stein1999}.
Because of these properties, $\sigma^2$ is called the {\it variance} parameter, 
$\vartheta$ the {\it range} parameter and $\nu$ the {\it smoothness} parameter.
%(the latter two are sometimes also called the scale and shape parameters, respectively).
 
In applications the measurements $z_i$ often contain measurement error,
in which case they are modeled as 
\begin{equation}
z_i \; = \; Z({\bf s}_i) \; + \; \epsilon_i, \quad\quad i=1,\ldots,n ,
\label{data-model}
\end{equation}
where $\epsilon_1,\ldots,\epsilon_n$ are i.i.d. with ${\rm N}(0,\tau^2)$ distribution and 
independent of $Z(\cdot)$; $\tau^2 \geq 0$ is called the {\it nugget} parameter. 
Although the findings in this work are likely to hold for other families of isotropic 
covariance functions, we focus on the Mat\'ern family where the covariance structure of the data 
is indexed by $\mbox{\boldmath $\eta$} := (\sigma^2, \tau^2, \vartheta, \nu)$ and
the smoothness parameter is considered unknown.

\section{Smoothness Parameter: Fix or Estimate?} 

\subsection{A Critical Look at a Geostatistical Practice}

As indicated in the Introduction, the claim that geostatistical data have little or no information 
about the smoothness of the random field is sometimes seen in geostatistical practice
\citep[page 866]{Bose2018}. 
In addition, \citet[page 113]{Diggle2007} state that
``\ldots when using the Mat\'ern correlation function, our experience has been that the
shape parameter $\kappa$ [$\nu$ in (\ref{eq:matern-cov})] is often poorly identified'',
and in page 114 they also state
``\ldots we have found that, for example, estimating all three parameters in the Mat\'ern model 
is very difficult because the parameters are poorly identified, leading to ridges or plateaus 
in the log--likelihood surface.''
As a workaround these authors suggest using likelihood evaluations
to select the smoothness parameter from a few candidate values, but this advice is often not followed.
Statements like the ones above are sometimes interpreted in geostatistical practice
as meaning that the data have little or no information about the smoothness parameter, 
prompting the practice of fixing the smoothness parameter at an arbitrary value.
More often than not, the exponential model ($\nu = 1/2$)  is used as the default model, 
and the other covariance parameters, variance, range and nugget are then estimated.
But this interpretation is overly simplistic and not granted in general,
since weak identifiability and an ill behavior of the likelihood surface both derive from the proposed model as a {\it whole}, and in general may not be
attributable to a single parameter, especially when the parameters are highly non--orthogonal.  

The practice of arbitrarily fixing the smoothness parameter assumes, perhaps implicitly,
at least one of the two tenets: 
(a) the data contain more information about the variance and range parameters than about 
the smoothness parameter and/or 
(b) the variance and range parameters are more important for spatial interpolation/prediction 
than the smoothness parameter.
However, there are theoretical and practical arguments that cast doubts about these tenets, 
at least for the Mat\'ern family.
It has been shown that, in the fixed--domain asymptotic framework,
$\sigma^2$ and $\vartheta$ {\it cannot} be consistently estimated when $d \leq 3$ \citep{Zhang2004}.
On the other hand, $\nu$ is what is called a {\it microergodic} parameter \citep{Stein1999},
so consistent estimation of $\nu$ is plausible. 
In fact, \cite{Wu2013}, \cite{Loh2015}, \cite{Wu2016} and \cite{Loh-etal-2021} 
have constructed, under some conditions on the design and the true smoothness, 
estimators of $\nu$ that are consistent under fixed--domain asymptotics. 
Hence, the aforementioned practice is in conflict with these results, which indicate
 that geostatistical data may contain substantial  information about the smoothness parameter,
at least under the conditions for which the above results hold.
Also, \citet[Theorem 3]{Kaufman2013} have shown that for any prediction location ${\bf s}_0$, 
the best linear predictor of $Z({\bf s}_0)$ based on a misspecified Mat\'ern model is (fixed--domain) 
asymptotically efficient when $\nu$ is correctly specified, 
regardless of the values of $\sigma^2$ and $\vartheta$.
The next section provides a practical example that casts doubt on the implicit two tenets mentioned above.
%aforementioned claim.

In this work we show that these tenets do not always hold, and investigate a more 
satisfactory practice to quantify information about covariance parameters based on 
the study of likelihood functions.
It is shown that the actual situation is much more nuanced than the above claim suggests, 
and that the information the data contain about the smoothness parameter depends critically on
aspects of the true model and observed data, in particular on the sampling design $\mathcal{S}_n$.

\subsection{A Telling Example}  \label{sec:telling-example}

\begin{figure}
\begin{center}
\psfig{figure=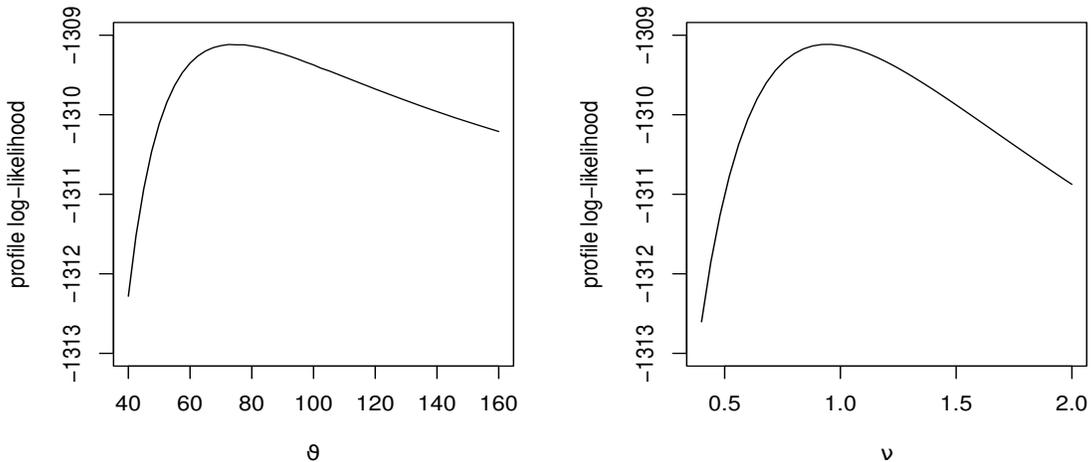, width=6in, height=3in}
\caption{Profile log--likelihood of $\vartheta$ (left) and $\nu$ (right) for the Swiss rainfall data.}
\label{sic-profile-loglik-vartheta-nu}
\end{center}
\end{figure}

In this section the rainfall data set analyzed in \citet[Section 5.4.7]{Diggle2007} is used 
to show that the claim of spatial data not being informative about the smoothness parameter 
does not hold. 
This data set is available from the {\tt R} package {\tt geoR}. 
The data consists of 467 measurements of daily rainfall collected in Switzerland on May 8, 1986, using an irregular sampling design where the coordinates of the sampling locations were measured in kilometers.
The model for the (square root transformed) data is of the form (\ref{data-model}), where 
$Z(\cdot)$ is a Gaussian random field with constant mean and the Mat\'ern covariance function 
(\ref{eq:matern-cov}). 
The maximum likelihood estimates of the covariance parameters are 
$\mbox{\boldmath $\hat{\eta}$} = 
(\hat{\sigma}^2, \hat{\tau}^2, \hat{\vartheta}, \hat{\nu}) = (105.09, 6.74, 73.42, 0.95)$.
The information about the covariance parameters \mbox{\boldmath $\eta$} may be quantified by 
inspecting the observed information matrix, $H(\mbox{\boldmath $\hat{\eta}$})$.
An approximation to this matrix is usually provided by the optimization algorithm, denoted as $\hat{H}$.
For the Swiss rainfall data, the inverse of this matrix is given by
\begin{equation}
\hat{H}^{-1} = \left(\begin{array}{cccc}
1720.183 & -5.633 & 838.860 & -2.431 \\
-5.633 & 2.817 & -17.850 & 0.474 \\
 838.860 & -17.850 & 626.629 & -5.223 \\
-2.431 & 0.474 & -5.223 & 0.105
\end{array}\right) .
\label{rain-obs-info-inv}
\end{equation}
For multiparameter cases, the information about a parameter may be measured by the inverse of 
the Cramer--Rao lower bound for the variance of unbiased estimators of that parameter. 
For $\nu$ say, this is estimated by $1/\hat{H}^{\nu \nu}$, with 
$\hat{H}^{\nu \nu}$ denoting the `$(\nu,\nu)$' diagonal element of $\hat{H}^{-1}$.
For the Swiss rainfall data $1/\hat{H}^{\nu \nu} = 9.524$, suggesting that these data contain substantial information about $\nu$.
Also, $1/\hat{H}^{\vartheta \vartheta} = 0.0016$.
Although this might suggest that these data are more informative about $\nu$ than 
about $\vartheta$, this is not necessarily the case since the information about $\vartheta$ 
depends on the (arbitrary) units used to measure distance (e.g., kilometers versus meters).
In Section \ref{sec:information-numerical-exploration} a re--scaling of the spatial coordinates 
is used so that
%reparametrization of the covariance model 
the information about the range and smoothness parameters can be more sensibly compared, and
this is used in Section \ref{sec:telling-example-continuation} to re--analyze this data set.
An equivalent visual approach to quantify information about the individual parameters 
is to inspect their profile log--likelihoods, ${\rm pl}_{1}(\vartheta)$ and ${\rm pl}_{2}(\nu)$
displayed in Figure \ref{sic-profile-loglik-vartheta-nu}.
Although the degree of ``peakness'' appears similar in both graphs,
the curvatures of these graphs at their maxima are quite different,
$-(\partial^2 / \partial \vartheta^2) {\rm pl}_{1}(\vartheta)\big|_{\vartheta=\hat{\vartheta}} 
\approx 1/\hat{H}^{\vartheta \vartheta} = 0.0016$ and 
$ -(\partial^2 / \partial \nu^2) {\rm pl}_{2}(\nu)\big|_{\nu=\hat{\nu}} 
\approx 1/\hat{H}^{\nu \nu} = 9.524$ \citep{Seber2003} 
%suggesting again that the information about the smoothness parameter is substantial
(the approximation is due to the use of the approximate rather than exact observed information matrix). 
\cite{Zhu-Zhang2006} provided another example of a data set that appears to contain 
substantial information about the smoothness parameter. 
The likelihood summaries reported in this section were obtained using the {\tt R} package {\tt georob} 
\citep{Papritz2021}.

\begin{figure}
\centering
\includegraphics[height=7cm, width=7.3cm]{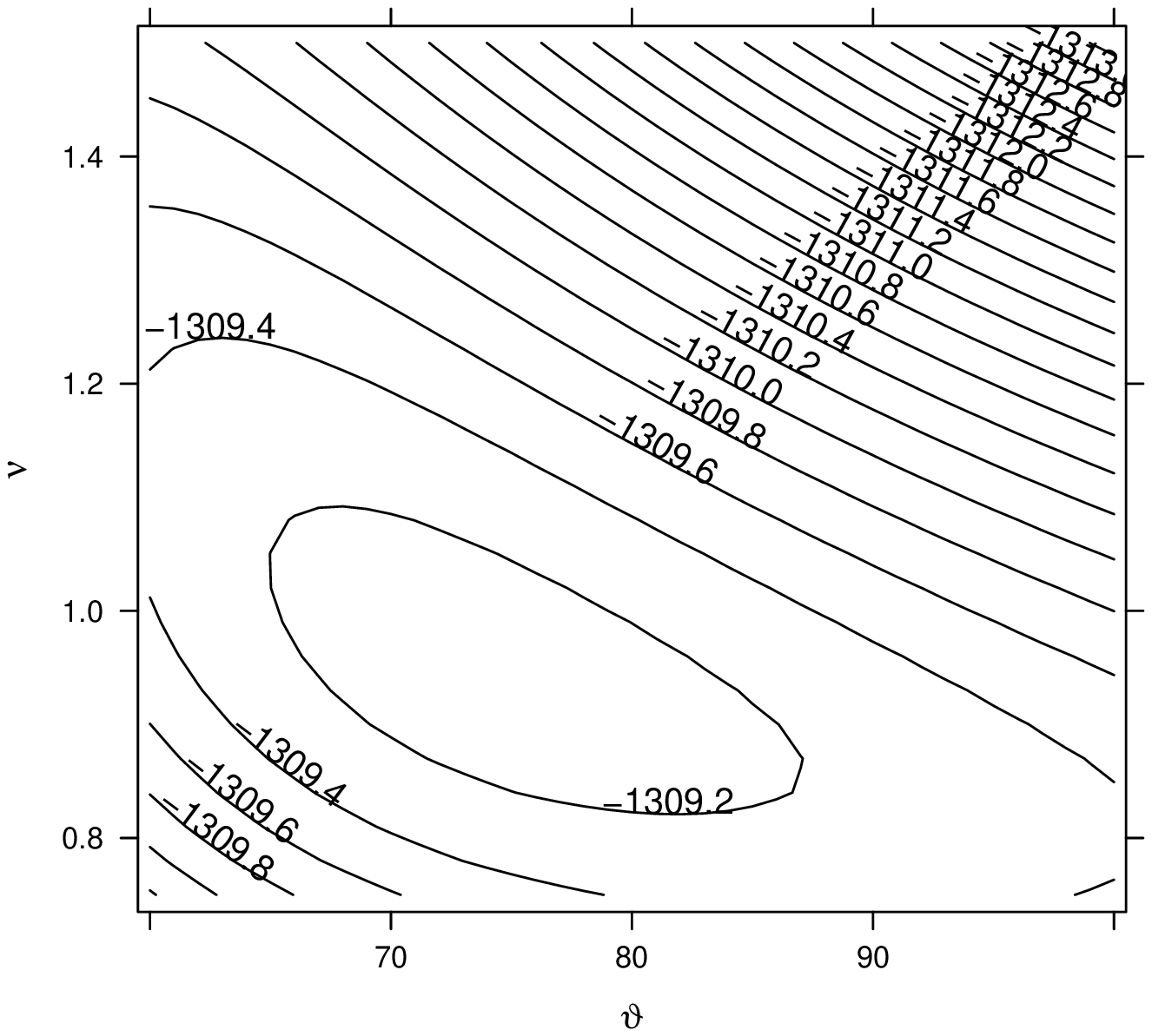} \  
\includegraphics[height=7cm, width=7.3cm]{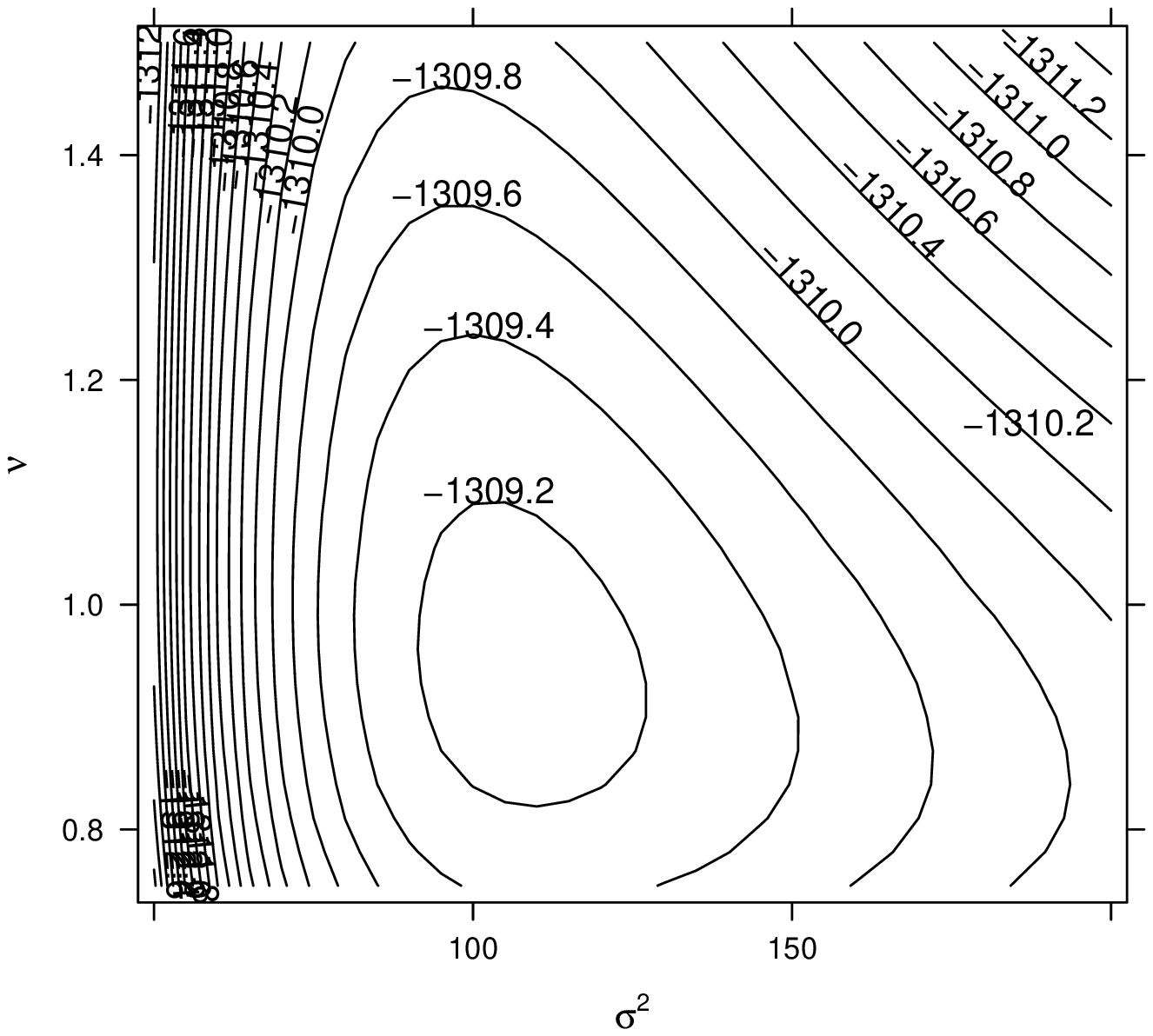} \\
\includegraphics[height=7cm, width=7.3cm]{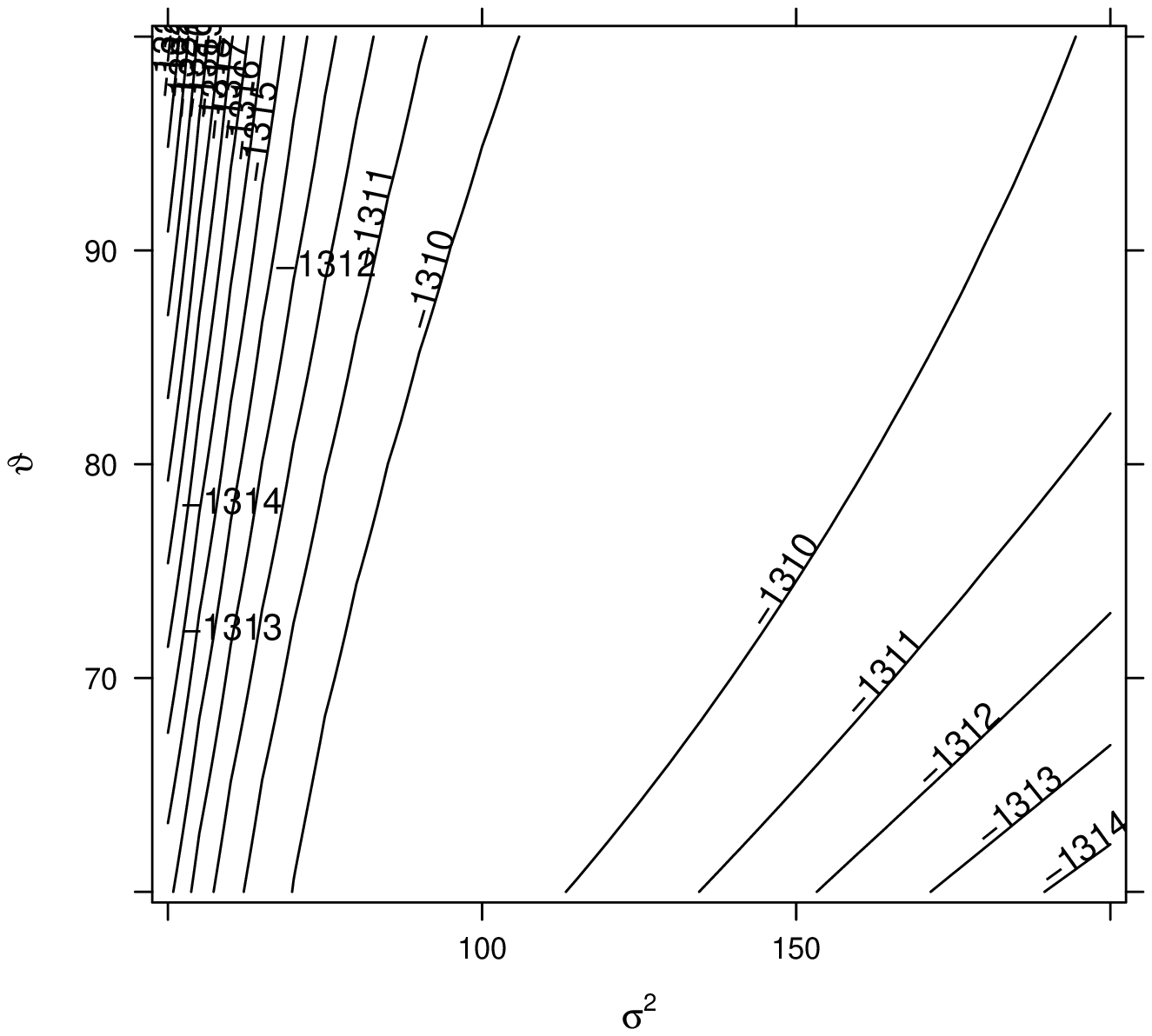} \
\includegraphics[height=7cm, width=7.3cm]{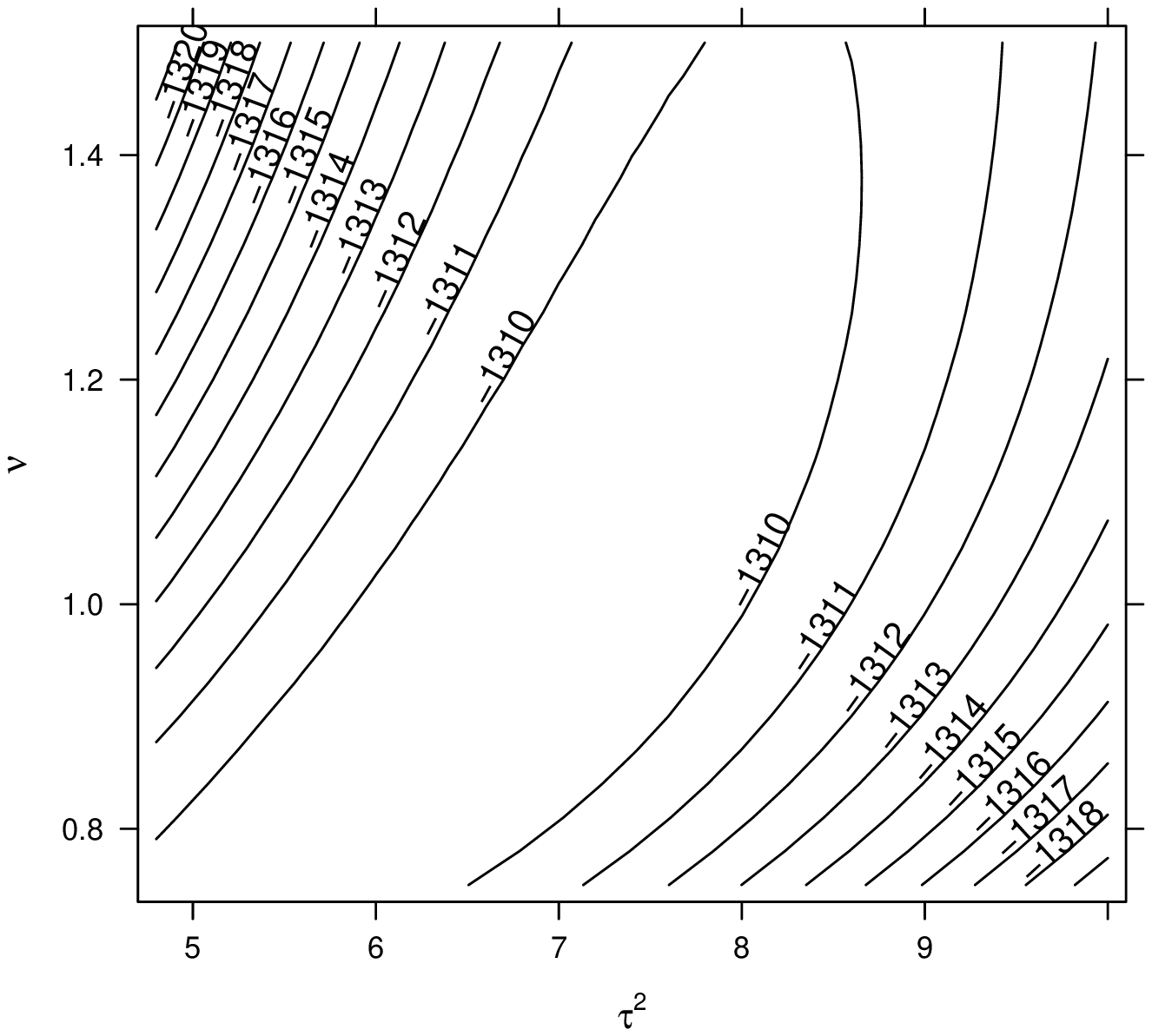} \\
\includegraphics[height=7cm, width=7.3cm]{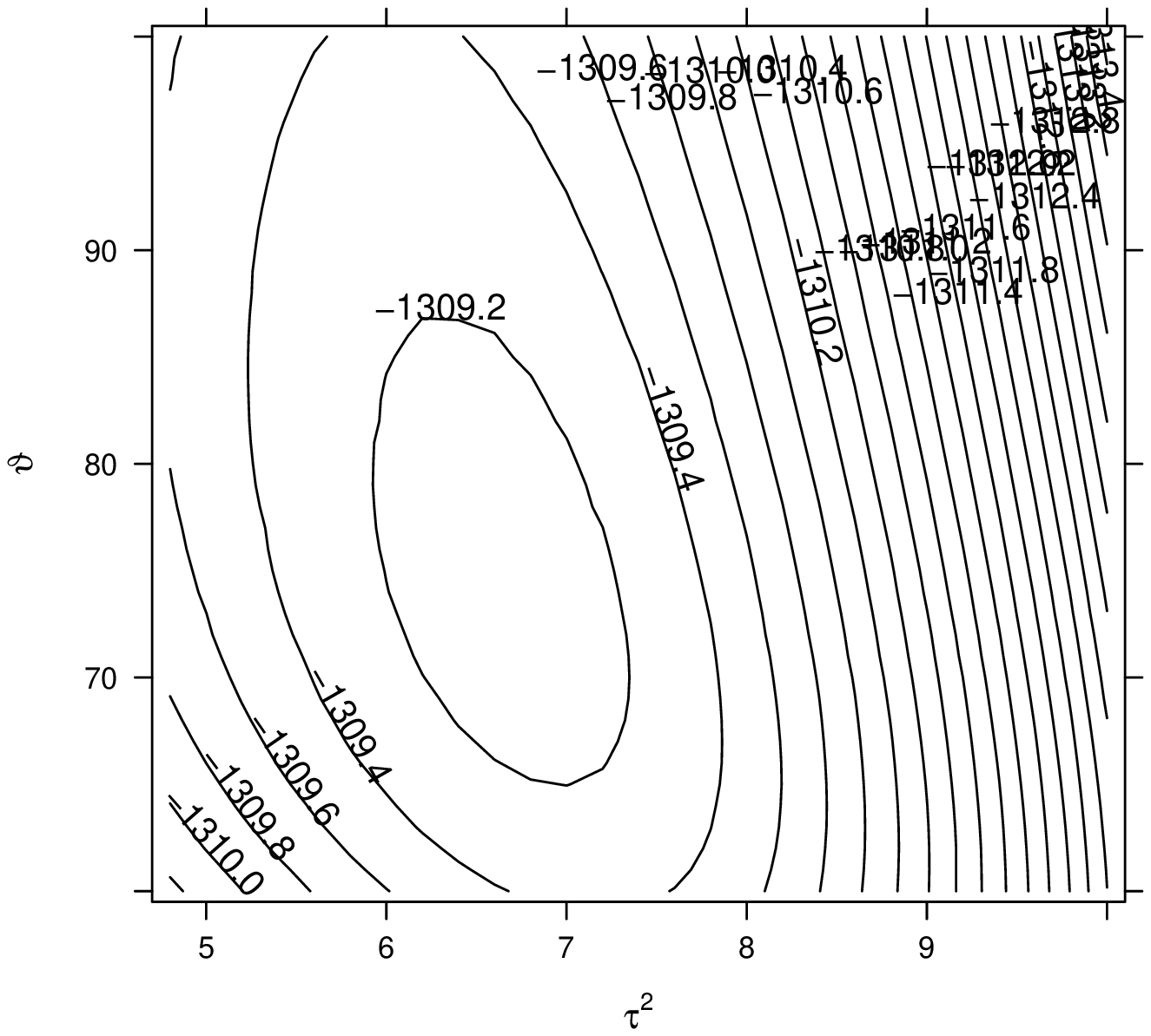} \ 
\includegraphics[height=7cm, width=7.3cm]{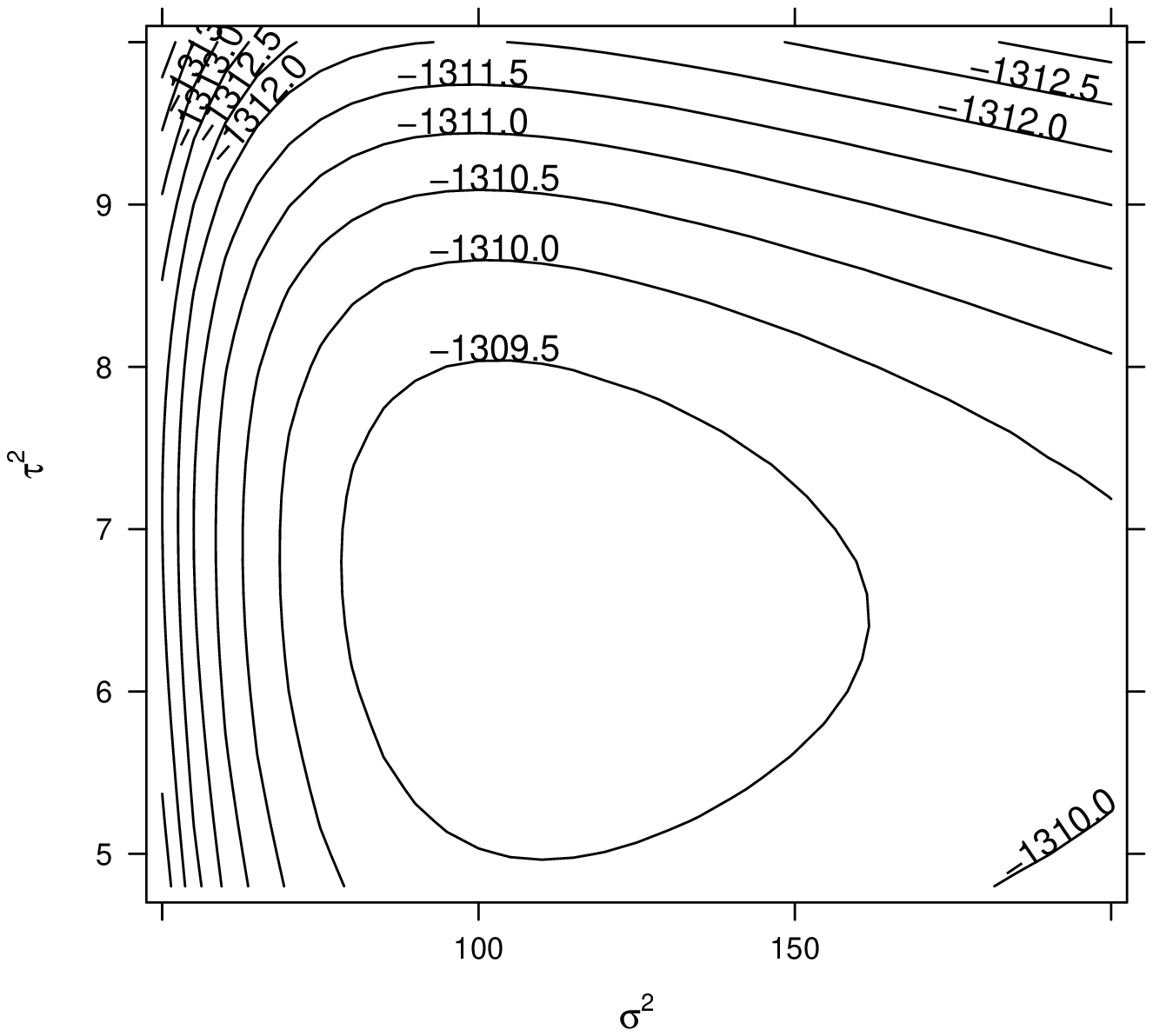}
\caption{Joint profile log--likelihoods of all pairs of parameters in $\mbox{\boldmath $\eta$}$ 
for the Swiss rainfall data.}
\label{sic-joint-profile-logliks}
\end{figure}

Since $\vartheta$ and $\nu$ are non--orthogonal and they are not the only model parameters, 
a more complete analysis also involves the inspection of joint profile log--likelihoods 
for all pairs of covariance parameters.
For the Swiss rainfall data, the contour plots of the joint profile log--likelihoods of 
$(\vartheta,\nu)$, $(\sigma^2,\nu)$, $(\sigma^2,\vartheta)$, $(\tau^2,\nu)$, $(\tau^2,\vartheta)$ 
and $(\sigma^2,\tau^2)$ are displayed in Figure \ref{sic-joint-profile-logliks}.
The joint profile log--likelihood of $(\vartheta,\nu)$ displays a plateau, showing that the MLEs of these parameters are highly interdependent.
%suggesting that there is a direction in the parameter space along which 
%linear combinations of $\vartheta$ and $\nu$ are poorly estimated.
A similar but more extreme behavior is displayed by the joint profile log--likelihoods of 
$(\sigma^2,\vartheta)$ and $(\tau^2,\nu)$.
On the other hand, the joint profile log--likelihood of $(\sigma^2,\nu)$ does not display a
plateau as its curvature at the maximum is much larger than that of the others, and
a similar behavior holds for the joint profile likelihood of $(\sigma^2,\tau^2)$. 
Together these plots suggest that the data might be more informative about $(\sigma^2,\nu)$
than about the other covariance parameters.

\subsection{Interplay Between Smoothness and Nugget}

It has been empirically observed by several authors and us, using both simulated and real data sets,
that the estimate of the nugget parameter is closely related to the assumed smoothness of 
the random field. 
Specifically, the more smooth the random field is assumed, the larger the estimate of the nugget; 
see \citet[Tables 5.1 and 5.2]{Diggle2007} and 
Figure \ref{sic-joint-profile-logliks} (middle right panel).
This is intuitively expected:
`discordant' observations collected at close by locations might be explained either as 
coming from a non--smooth random field model or due to the presence of measurement error.
If the smoothness were to be fixed at a value that is higher than the one supported by the data, 
an overestimation of the nugget would result to compensate.
The opposite effect is expected when the smoothness is fixed at a value that is too low.
As an illustration, if when fitting the Swiss rainfall data the smoothness parameter 
is fixed in advance at the values $\nu=0.5$, $1$ and $1.5$, the corresponding estimates of 
the nugget are $\hat{\tau}^2 = 2.48$, $6.90$ and $8.17$, respectively.

From the above follows that arbitrarily fixing the smoothness parameter may result in a substantial 
misspecification of the nugget parameter, which has consequences for spatial prediction.
When the data contain measurement error, the so--called nugget--to--sill ratio, 
$\tau^2 / (\sigma^2 + \tau^2)$, controls the amount of `smoothing' (as opposed to interpolation) 
that is carried out on the data for spatial prediction.
For the Swiss rainfall data, the estimated nugget--to--sill ratios are $0.02$, $0.06$ and $0.08$
when the smoothness parameter is fixed at one of the aforementioned values.
Estimating the smoothness parameter from the data, rather than fixing it, 
may avoid an undue dependence of the estimated nugget (and  nugget--to--sill ratio)
on a possibly grossly misspecified smoothness.
%Another question I plan to address is to determine which of the two is more detrimental for 
%predictive inference, fixing the smoothness parameter at a value that is too high or too low.

\section{Quantifying Information About Covariance Parameters}

In this section we carry out a numerical exploration to uncover the extent to which 
the sampling design and true model affect the information content the data have about 
covariance parameters.
We consider several sampling designs in the plane ($d=2$) that are commonly used in 
geostatistical practice, 
and describe a new way to efficiently compute the Fisher information matrix for the Mat\'ern model.
%$\mathcal{S}_n$ (e.g., average distance between nearest neighbors) 
As in the example in Section \ref{sec:telling-example}, the information content about each  
covariance parameter is measured by the vector
\begin{equation}
{\rm Inf}(\mbox{\boldmath $\eta$},\mathcal{S}_n) :=
\Bigg( \frac{1}{I(\mbox{\boldmath $\eta$},\mathcal{S}_n)^{\sigma^2 \sigma^2}},
\frac{1}{I(\mbox{\boldmath $\eta$},\mathcal{S}_n)^{\tau^2 \tau^2}},
\frac{1}{I(\mbox{\boldmath $\eta$},\mathcal{S}_n)^{\vartheta \vartheta}},
\frac{1}{I(\mbox{\boldmath $\eta$},\mathcal{S}_n)^{\nu \nu}} \Bigg) ,
\label{inf-vec}
\end{equation}
where $I(\mbox{\boldmath $\eta$},\mathcal{S}_n)^{\sigma^2 \sigma^2}$,
$I(\mbox{\boldmath $\eta$},\mathcal{S}_n)^{\tau^2 \tau^2}$,
$I(\mbox{\boldmath $\eta$},\mathcal{S}_n)^{\vartheta \vartheta}$ and
$I(\mbox{\boldmath $\eta$},\mathcal{S}_n)^{\nu \nu}$ are, respectively, the first, second, 
third and fourth diagonal elements of $I(\mbox{\boldmath $\eta$},\mathcal{S}_n)^{-1}$,
and $I(\mbox{\boldmath $\eta$},\mathcal{S}_n)$ is the Fisher information matrix based on
the sampling design $\mathcal{S}_n$ when the true covariance parameter is $\bfeta$.

\subsection{Sampling Designs}  \label{sec:designs}

It is well established in the geostatistical literature that the sampling design exerts
a substantial effect on the properties of parameter estimators and predictors.
Designs that are most favorable for parameter estimation are quite different from those that 
are most favorable for spatial prediction, under the assumption that the parameters are known 
\citep{Zhu-Stein2005,Zhu-Zhang2006,Zimmerman2006}. % Irvine, Gitelman and Hoeting, 2007).
Designs that are optimal for covariance function estimation are irregular, in the sense that 
they include a substantial fraction of clustered or closely spaced sampling locations, while 
designs that are optimal for spatial prediction tend to be regular or `space filling'.
Because of this, some `hybrid' designs have been proposed that supplement a set of
regularly spaced sampling locations with a set of closely spaced sampling locations,
with the intent of balancing the (conflicting) requirements for adequate covariance estimation 
and spatial prediction.
See \cite{Diggle-Lophaven2006}, \cite{Zhu-Stein2006} and \cite{Zimmerman2006} 
for examples of hybrid designs.

To assess the effect of the sampling design $\mathcal{S}_n$ on the amount of information 
the data have about covariance parameters, we consider the four types of
sampling designs described below, which are commonly used in practice.
For concreteness, in the exploration in Section \ref{sec:information-numerical-exploration} 
we define these sampling designs for the region $\mathcal{D} = [0,1] \times [0,1]$ and sample size 
$n=225$ or $226$, but they can be equally defined for other regions and sample sizes.

\smallskip

{\it Regular}.
This is a deterministic design where the sampling locations form a $15 \times 15$ regular lattice 
in $\mathcal{D}$; the distance between neighboring sampling locations is $r_{\min} = 1/15 \approx 0.066$.
This design is preferred when the goal is spatial prediction with a known model.

\smallskip

{\it Random}.
This is a random design where the sampling locations are a random sample of size $225$ from 
the unif$\big((0,1)^2\big)$ distribution.
This design is preferred when the goal is to estimate the covariance parameters.

\smallskip

{\it Bachoc}.
This is a class of random designs proposed by \cite{Bachoc2014} and is constructed as follows. 
For $n = n_1 n_2$, with $n_1, n_2 \in {\mathbb N}$, let ${\bf v}_1,\ldots,{\bf v}_n$ be a set of 
points in $\mathcal{D}$ that form a regular lattice with distance $\Delta > 0$ 
between neighboring ${\bf v}_i$s, and ${\bf X}_1,\ldots,{\bf X}_n$ be a set of i.i.d. 
random vectors with a distribution symmetric about ${\bf 0}_2$ and with support contained 
in $(-\Delta,\Delta)^2$. 
The sampling design $\mathcal{S}_n$ is assumed to be a realization of the set
$\{{\bf v}_1 + \epsilon {\bf X}_1,\ldots,{\bf v}_n + \epsilon {\bf X}_n\}$
for some $\epsilon \in [0,1/2)$.
By varying the tuning constant $\epsilon$ this scheme can generate a continuum of designs
that range from regular ($\epsilon = 0$) to moderately irregular ($\epsilon \approx 1/2$).
We assume in Section \ref{sec:information-numerical-exploration} that
$n_1 = n_2 = 15$, $\Delta = 1/15$, $\epsilon = 0.4$ and ${\bf X}_i \sim {\rm unif}\big( (-\Delta,\Delta)^2\big)$.

\begin{figure}[t!]
\begin{center}
\psfig{figure=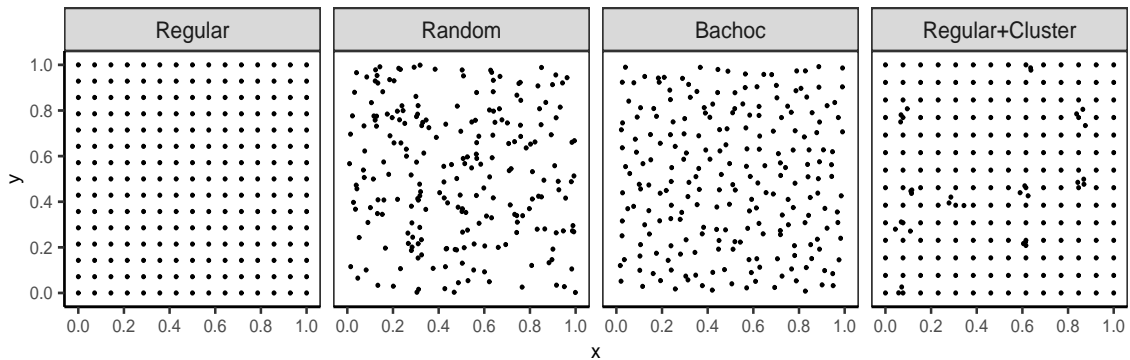, width=6in, height=2in} 
\caption{Examples of four different sampling design types in $\mathcal{D} = [0, 1] \times [0, 1]$.}
\label{designs}
\end{center}
\end{figure}

\smallskip

{\it Regular+Cluster}.
This is a class of random designs consisting of a set of points in a regular design supplemented with 
several sets of highly clustered points. It is constructed in two steps as follows.
First, $n_1 \times n_1$ sampling locations are selected that form a regular lattice in  $\mathcal{D}$.
Second, $nc$ points are selected at random (without replacement) from the $n_1^2$ points in the
first step to serve as cluster centers.
Then for each cluster center, ${\bf v}_i$ say, a set of $ppc - 1$ points is generated independently 
and uniformly distributed around the cluster center as ${\bf v}_i + {\bf w}_{ij}$, with 
${\bf w}_{ij} \sim {\rm unif}((-\epsilon,\epsilon)^2)$ for some $\epsilon > 0$.
This is a kind of `hybrid' design with $n = n_1^2 + nc(ppc -1)$ sampling locations 
that seeks to balance the requirements for adequate covariance estimation and spatial prediction.
We assume in Section \ref{sec:information-numerical-exploration} that 
$n_1 = 14$, $nc = 10$, $ppc = 4$ and $\epsilon = 0.04$.

\smallskip

The above sampling designs are illustrated in Figure \ref{designs}, where a generic 
sampling location $\bf s$ in a design has Cartesian coordinates $(x,y)$.

\subsection{Computation of Fisher Information Matrix: Mat\'ern Model}  \label{sec:fisher-info}

The results reported in Section \ref{sec:telling-example} used an approximation of the 
{\it observed} information matrix based on difference quotients, as provided by
the output of the optimization algorithm. 
But this can be a poor approximation to the {\it Fisher} information matrix.
%see Section \ref{sec:telling-example-continuation}.
A more reliable alternative would be to use an explicit expression for the Fisher information matrix, 
when this is available.
Let $\mbox{\boldmath $\eta$} = (\eta_1, \eta_2, \eta_3, \eta_4) 
= (\sigma^2, \tau^2, \vartheta, \nu)$ be the covariance parameters.
For Gaussian random fields the Fisher information matrix of $\mbox{\boldmath $\eta$}$
based on the data model in (\ref{data-model}) is the $4 \times 4$ matrix 
$I(\mbox{\boldmath $\eta$},\mathcal{S}_n)$ with entries 
\citep{Cressie1993,Stein1999}
\begin{equation}
I(\mbox{\boldmath $\eta$},\mathcal{S}_n)_{ij} = \frac{1}{2} 
{\rm tr}\big( \Psi^{-1}(\mbox{\boldmath $\eta$},\mathcal{S}_n) 
\Psi_{i}(\mbox{\boldmath $\eta$},\mathcal{S}_n) \Psi^{-1}(\mbox{\boldmath $\eta$},\mathcal{S}_n) 
\Psi_{j}(\mbox{\boldmath $\eta$},\mathcal{S}_n) \big) ,
\label{fisher-information}
\end{equation}
where $\Psi(\mbox{\boldmath $\eta$},\mathcal{S}_n) := 
\sigma^2 \boldsymbol{\Sigma}_{\boldsymbol{\vartheta}} + \tau^2 \boldsymbol{I}_n$,
$\boldsymbol{\Sigma}_{\boldsymbol{\vartheta}}$  is the $n \times n$ matrix with entries
$(\boldsymbol{\Sigma}_{\boldsymbol{\vartheta}})_{ij} = 
K_{\boldsymbol{\vartheta}}(\| {\bf s}_i - {\bf s}_j\|)$, $\boldsymbol{\vartheta} = (\vartheta, \nu)$,
$\boldsymbol{I}_n$ is the $n \times n$ identity matrix and 
$\Psi_{i}(\mbox{\boldmath $\eta$},\mathcal{S}_n) := 
\frac{\partial}{\partial \eta_i}\Psi(\mbox{\boldmath $\eta$},\mathcal{S}_n)$,
%$\frac{\partial}{\partial \eta_i} \Psi(\mbox{\boldmath $\eta$},\mathcal{S}_n)$
where these derivatives are computed entry--wise.
The above matrices do not depend on $p$ nor the regression parameters $\bfbeta$. 
%and their dependence on the sampling design is not made explicit to avoid making the notation overly cumbersome.
For the Mat\'ern model, the computation of some of the entries of $I(\mbox{\boldmath $\eta$},\mathcal{S}_n)$
requires computing derivatives of the Bessel function $\mathcal{K}_{\nu}(x)$, 
both with respect to $x$ and $\nu$.
%For the former derivative it holds that (Gradshteyn and Ryzhik, 2000, 8.486--11)
%\begin{equation}
%\frac{\partial}{\partial x} \mathcal{K}_{\nu}(x)
%= -\frac{1}{2}\big( \mathcal{K}_{\nu - 1}(x) + \mathcal{K}_{\nu + 1}(x) \big) ,
%\label{derv-K-x}
%\end{equation}
For the former derivative it holds that \citep[8.486--12]{Gradshteyn2000}
\begin{equation}
\frac{\partial}{\partial x} \mathcal{K}_{\nu}(x)
= -\Big( \mathcal{K}_{\nu - 1}(x) + \frac{\nu}{x}\mathcal{K}_{\nu}(x) \Big) , \quad\quad x \neq 0 ,
\label{deriv-K-x}
\end{equation}
so this can be computed exactly from the code that computes this Bessel function
(e.g., the {\tt R} function {\tt besselK}).
For the latter derivative and when $\nu = m$ is a non--negative integer, it holds that \citep[8.486(1)--9]{Gradshteyn2000}
%(Gradshteyn and Ryzhik, 2000, 8.486(1)--9)
\begin{equation}
\frac{\partial}{\partial \nu} \mathcal{K}_{\nu}(x)\Big|_{\nu=m} = \
\frac{m!}{2} \sum_{j=0}^{m-1} \frac{\big(x/2\big)^{j - m} \mathcal{K}_{j}(x)}{j! (m - j)},
\label{deriv-K-nu-int}
\end{equation}
so again this can be computed from the code that computes this Bessel function.
%this relation was used in the numerical exploration in Stein (1999, Section 6.6).
On the other hand, it seems to be few algorithms available to compute this derivative 
when $\nu$ is not a non--negative integer.
An approach to compute this derivative for an arbitrary value of $\nu$ is to use 
a representation of this Bessel function that is amenable to exact differentiation. 
One such representation is given by \citep[8.432--1]{Gradshteyn2000}
%(Gradshteyn and Ryzhik, 2000, 8.432--1)
\[
\mathcal{K}_{\nu}(x) = \int_{0}^{\infty} e^{-x \cosh(t)} \cosh(\nu t) dt , 
\]
from which it follows that %, after interchanging of the derivative and integral, 
\begin{equation}
\frac{\partial}{\partial \nu} \mathcal{K}_{\nu}(x) = 
\int_{0}^{\infty} t e^{-x \cosh(t)} \sinh(\nu t) dt .
\label{deriv-K-nu}
\end{equation}
%Inspection of the above integrand for any $x, \nu > 0$ reveals that it is a 
%smooth function that converges to zero very fast as $t$ approaches zero or infinity.
%Initial explorations of the approximation of the latter integral, either by an
%adaptive numerical quadrature algorithm (e.g., the {\tt R} function {\tt integrate}) or
%by the Gauss--Laguerre quadrature algorithm, shows that the approximation by any of those 
%algorithms is very fast and accurate. For a variety of test combinations of $x > 0$ and $\nu=m$ 
%a non--negative integer, this approach produces an approximation to
%$\frac{\partial}{\partial \nu} \mathcal{K}_{\nu}(x)$ that is indistinguishable, 
%for all practical purposes, from the exact value in (\ref{derv-K-nu}).

Using (\ref{deriv-K-x}) and (\ref{deriv-K-nu}) we have that for any 
$\boldsymbol{\vartheta} \in (0, \infty)^2$ and $r \geq 0$  
%\begin{equation}
%\frac{\partial}{\partial \vartheta} K_{\boldsymbol{\vartheta}}(r) =
%\frac{2(\sqrt{\nu}r)^{\nu}}{\Gamma(\nu)\vartheta^{\nu + 1}} \left( \frac{\sqrt{\nu}r}{\vartheta}
%\Big[ \mathcal{K}_{\nu - 1}\Big( \frac{2\sqrt{\nu}}{\vartheta} r \Big)  
%+ 	\mathcal{K}_{\nu + 1}\Big( \frac{2\sqrt{\nu}}{\vartheta} r \Big) \Big] 
%- \nu \mathcal{K}_{\nu}\Big( \frac{2\sqrt{\nu}}{\vartheta} r \Big)	\right) ,
%\label{deriv-corr-mat-vartheta}
%\end{equation}
\begin{equation}
\frac{\partial}{\partial \vartheta} K_{\boldsymbol{\vartheta}}(r) =
\frac{4 \nu^{\frac{\nu + 1}{2}} r^{\nu+1}}{\Gamma(\nu)\vartheta^{\nu + 2}} 
\mathcal{K}_{\nu - 1}\Big( \frac{2\sqrt{\nu}}{\vartheta} r \Big)  ,
\label{deriv-corr-mat-vartheta}
\end{equation}
where $K_{\boldsymbol{\vartheta}}(r)$ is the Mat\'ern correlation function 
defined in (\ref{eq:matern-cov}) and
%\begin{align}
%\frac{\partial}{\partial \nu} K_{\boldsymbol{\vartheta}}(r) & =
%\left(\frac{1}{2} + \log\left(\frac{\sqrt{\nu}}{\vartheta} r\right) - \psi(\nu)\right)
%	K_{\boldsymbol{\vartheta}}(r)  
%\label{deriv-corr-mat-nu} \\
%& \ \ \ \ + \ h(\nu) \int_{0}^{\infty} 
%\left[ t\sinh(\nu t) - \frac{r}{\vartheta \sqrt{\nu}} \cosh(\nu t)\cosh(t)
%\right] \exp\left(-\frac{2r\sqrt{\nu}}{\vartheta}\cosh(t)\right) dt , \nonumber
%\end{align}
\begin{align} \label{deriv-corr-mat-nu}
\frac{\partial}{\partial \nu} K_{\boldsymbol{\vartheta}}(r) & =
\left(\log\left(\frac{\sqrt{\nu}}{\vartheta} r\right) - \psi(\nu)\right)
K_{\boldsymbol{\vartheta}}(r)  \\
& \ \ \ \ -  h(\nu) \left( \frac{r}{\vartheta \sqrt{\nu}} 
\mathcal{K}_{\nu - 1}\left( \frac{2\sqrt{\nu}}{\vartheta} r \right)
- \int_{0}^{\infty} t\sinh(\nu t) \exp\left(-\frac{2r\sqrt{\nu}}{\vartheta}\cosh(t)\right) dt \right),
\nonumber
\end{align}
where $ \psi(\nu)$ is the digamma function and
\[
h(\nu) := \frac{2}{\Gamma(\nu)}
\left( \frac{\sqrt{\nu}}{\vartheta} r \right)^{\nu} ; 
%\left(= \frac{K_{\boldsymbol{\vartheta}}(r)}{\mathcal{K}_{\nu}\left( \frac{2\sqrt{\nu}}{\vartheta} r \right)} 
%\right) ;
\]
the derivations of the above identities are given in the Appendix.
For any $\nu > 0$ and $w := 2r\sqrt{\nu}/\vartheta >0$, 
inspection of the integrand of the integral in (\ref{deriv-corr-mat-nu}) reveals that 
it is a positive, smooth and unimodal function that converges to zero very fast as 
$t$ approaches zero or infinity.
Approximation of this integral (e.g., using the {\tt R} function {\tt integrate}) is relatively
fast and accurate for the purpose at hand (but see Section \ref{conclusions-discussion}).
For instance, for a variety of test combinations of $w > 0$ and $\nu=m$ a non--negative integer, 
it was found that these approximations to $({\partial}/{\partial \nu}) K_{\boldsymbol{\vartheta}}(r)$ 
are indistinguishable from the exact values obtained by using (\ref{deriv-K-nu-int}).

\subsection{Numerical Exploration of Information Patterns}  \label{sec:information-numerical-exploration}

In this subsection we explore the patterns of variation of the information defined 
in (\ref{inf-vec}) under various model settings and sampling designs.
Consider collecting data in the region of the plane $\mathcal{D} = [0,1] \times [0,1]$
that follow model (\ref{data-model}) with $C_{\boldsymbol{\theta}}(r)$ the Mat\'ern covariance function 
in (\ref{eq:matern-cov}). Without loss of generality we assume $p=1$ and $\beta = 0$.
For the sampling designs $\mathcal{S}_n$ defined in Section \ref{sec:designs} and a set of
representative covariance parameters $\bfeta$, we explore the variation of 
the last two components of the information vector 
${\rm Inf}(\mbox{\boldmath $\eta$},\mathcal{S}_n)$ in (\ref{inf-vec})
as well as that of a particular microergodic parameter;
a similar but more limited exploration was reported in \cite[Section 6.6]{Stein1999} for processes in the line.
For the rest of this section the spatial coordinates are re--scaled.
Specifically, we use new coordinates defined as 
$\tilde{\bf s} = (\tilde{x}, \tilde{y}) := {\bf s} /r_{\rm max}$,
where $r_{\rm max} := \max\{||{\bf s} - {\bf u}|| : {\bf s}, {\bf u} \in \mathcal{D}\}$;
$\tilde{\bf s} := (x,y)/\sqrt{2}$ for the aforementioned region $\mathcal{D}$.
Then, the re--scaled coordinates are unitless and invariant to the units of the original coordinates,
and consequently so is the range parameter.
The purpose of the re--scaling is to be able to compare more sensibly the information about 
the range and smoothness parameters, so the former may serve as a reference to judge when 
the latter is substantial.
Although other re--scalings are possible, all produce the same effect, namely, 
making the range parameter unitless while changing little the information about the other
covariance parameters.

Except for the regular design, the other three designs are random, and so is 
${\rm Inf}(\mbox{\boldmath $\eta$},\mathcal{S}_n)$.
Nevertheless, numerical inspection of ${\rm Inf}(\mbox{\boldmath $\eta$},\mathcal{S}_n)$
reveals that, when the covariance parameters are kept fixed, the entries of this vector 
vary very little over different realizations of the same design type (of the same size), and 
the same holds for their ordering.
Because of this, we ignore the stochastic nature of ${\rm Inf}(\mbox{\boldmath $\eta$},\mathcal{S}_n)$ 
and investigate the patterns of variation for a single realization of each of the considered sampling designs.
As for the covariance parameters, we fix $\sigma^2 = 1$ and $\tau^2 = 0.2$ and explore
the variations of the last two components of 
${\rm Inf}(\mbox{\boldmath $\eta$},\mathcal{S}_n)$, denoted 
${\rm Inf}(\mbox{\boldmath $\eta$},\mathcal{S}_n)_{3}$ 
and ${\rm Inf}(\mbox{\boldmath $\eta$},\mathcal{S}_n)_{4}$,
as a function of the correlation parameters in a grid of points 
$(\vartheta,\nu)$ in $[0.05, 0.65] \times [0.1, 1.5]$. 
This set of correlation parameters includes range parameters that are practically relevant for
the region $\mathcal{D}$ and smoothness parameters that are commonly found in geostatistical practice.

\medskip

{\bf Information About the Range Parameter}.
We first investigate the variation of the information about the range parameter $\vartheta$,
both as a function of $\vartheta$ and as a function of $\nu$.

\begin{figure}[t!]
\begin{center}
\psfig{figure=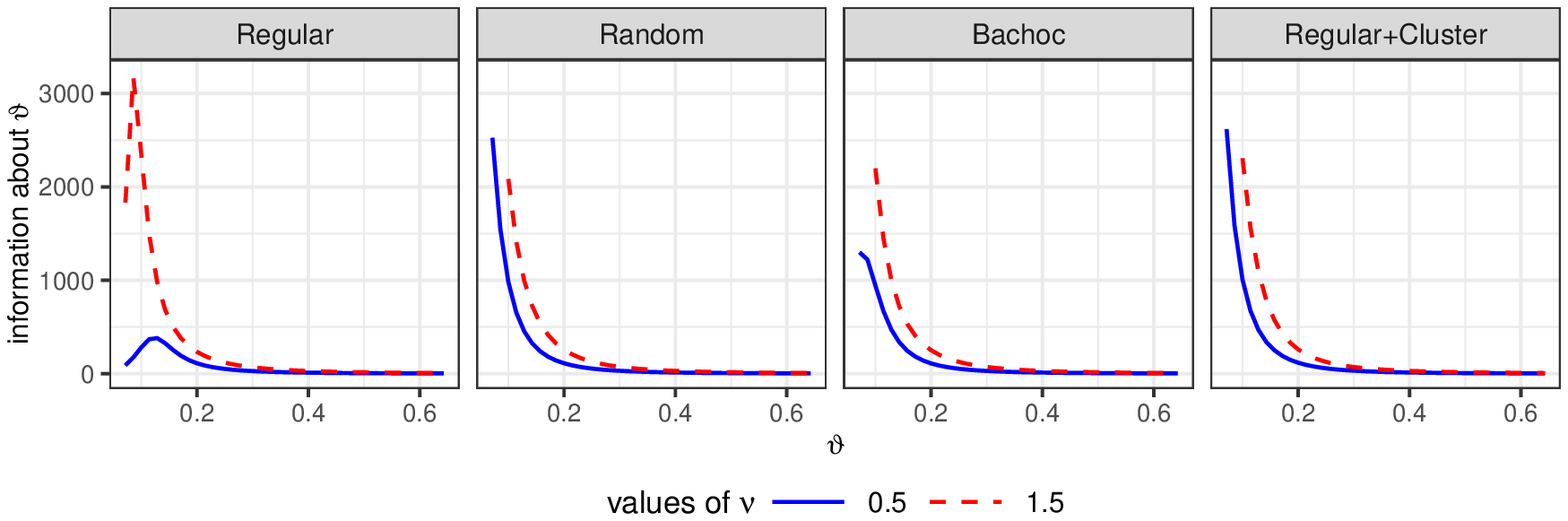, width=6in, height=2.5in} \\
\vspace{0.5 cm}
\psfig{figure=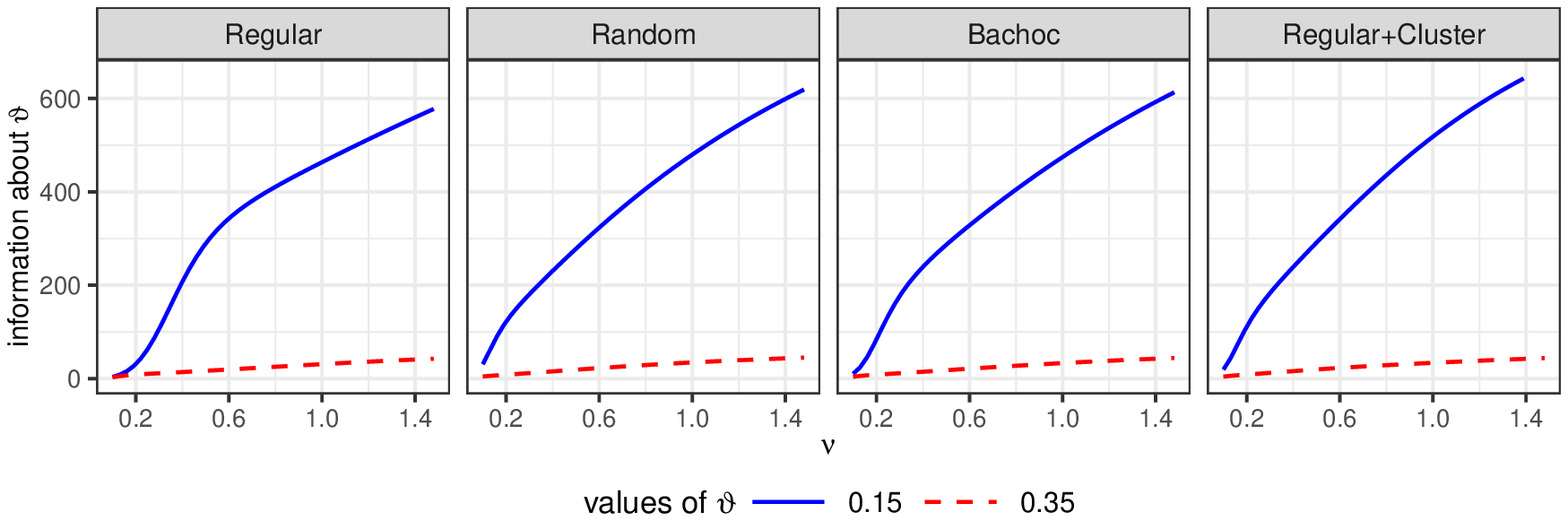, width=6in, height=2.5in}
\end{center}
\caption{Plots of the information about the range parameter $\vartheta$, ${\rm Inf}(\mbox{\boldmath $\eta$},\mathcal{S}_n)_{3}$,
%$1/I(\mbox{\boldmath $\eta$},\mathcal{S}_n)^{\vartheta \vartheta}$,
for different sampling designs, $\sigma^2 = 1$ and $\tau^2 = 0.2$. 
Top: Information about $\vartheta$ as a function of $\vartheta$ for two values of $\nu$. 
Bottom: Information about $\vartheta$ as a function of $\nu$ for two values of $\vartheta$.} 
\label{fig:info-rho}
\end{figure}

Figure \ref{fig:info-rho} (top panels) displays plots of 
${\rm Inf}(\mbox{\boldmath $\eta$},\mathcal{S}_n)_{3}$
%$1/I(\mbox{\boldmath $\eta$},\mathcal{S}_n)^{\vartheta \vartheta}$
as a function of $\vartheta$ when $\nu = 0.5$ and $1.5$ for the four sampling designs.
For both values of the smoothness parameter and all the designs but the Regular, 
the information about $\vartheta$ decreases monotonically when $\vartheta$ increases, which is expected 
as the `effective sample size' decreases when the strength of spatial correlation increases.
The Regular design departs slightly from this pattern for small values of $\vartheta$.
%the information about $\vartheta$ increases  up to a point, after which it monotonically decreases.
Also, regardless of the design and the range parameter, the information about $\vartheta$ is larger 
when $\nu=1.5$ than when $\nu=0.5$.
A possible explanation for this pattern (and a similar one described in the paragraph below)
is that realizations of smooth processes are less oscillatory than those of non--smooth processes,
so the strength of correlation can be better assessed based on data from the former.
In addition, the information about $\vartheta$ is very small when the spatial correlation is strong and,  
for most models, the information about the range parameter is quite similar across all the designs.

Figure \ref{fig:info-rho} (bottom panels) displays plots of ${\rm Inf}(\mbox{\boldmath $\eta$},\mathcal{S}_n)_{3}$ 
%$1/I(\mbox{\boldmath $\eta$},\mathcal{S}_n)^{\vartheta \vartheta}$
as a function of $\nu$ when $\vartheta = 0.2$ and $0.5$ for the four sampling designs.
For both values of the range parameter and all the designs, 
the information about $\vartheta$ increases monotonically when $\nu$ increases.
The rate of increase is fast when $\vartheta = 0.2$ while it is very slow when $\vartheta = 0.5$.
Also, regardless of the design and the smoothness parameter, the information about $\vartheta$
is substantially larger when $\vartheta=0.2$ than when $\vartheta=0.5$.
In addition, the information about $\vartheta$ is very large when the spatial correlation is weak and 
the process is smooth, and once again the information about the range parameter displays little
sensitivity to the different designs.
Overall, the plots in Figure \ref{fig:info-rho} support the common empirical finding 
that inference about range parameters is difficult when the data are highly correlated, but 
they also suggest that this challenge abates somewhat for smooth processes.

\begin{figure}[t!]
\begin{center}
\psfig{figure=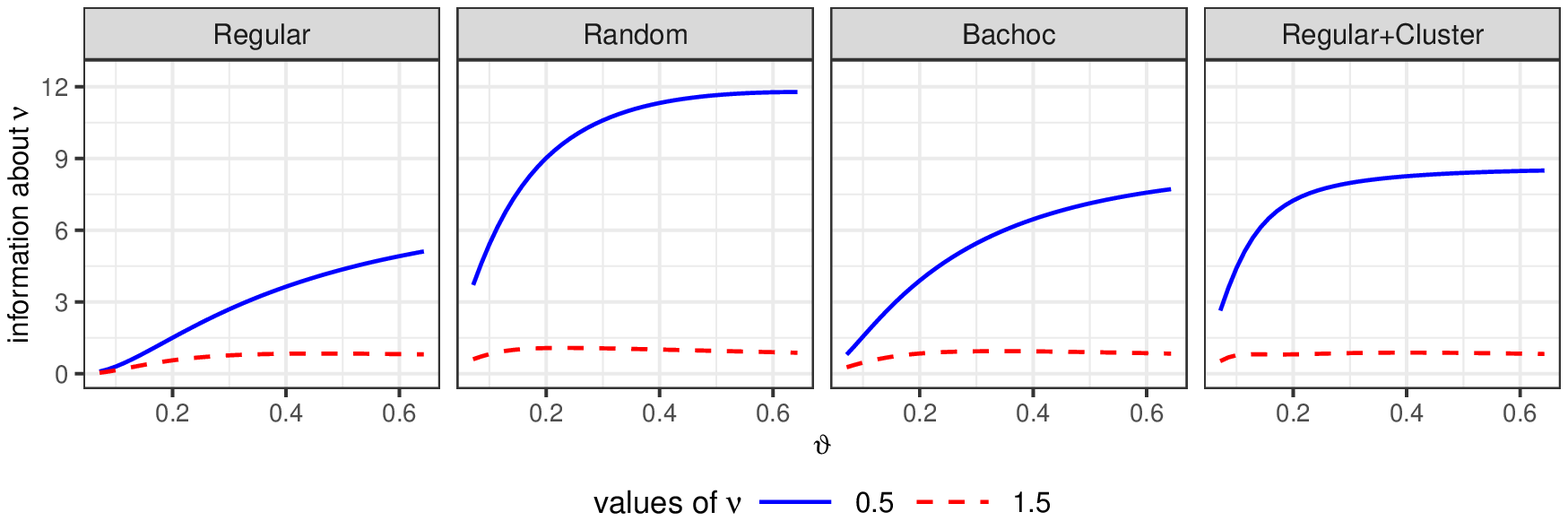, width=6in, height=2.5in}
\vspace{0.5 cm}
\psfig{figure=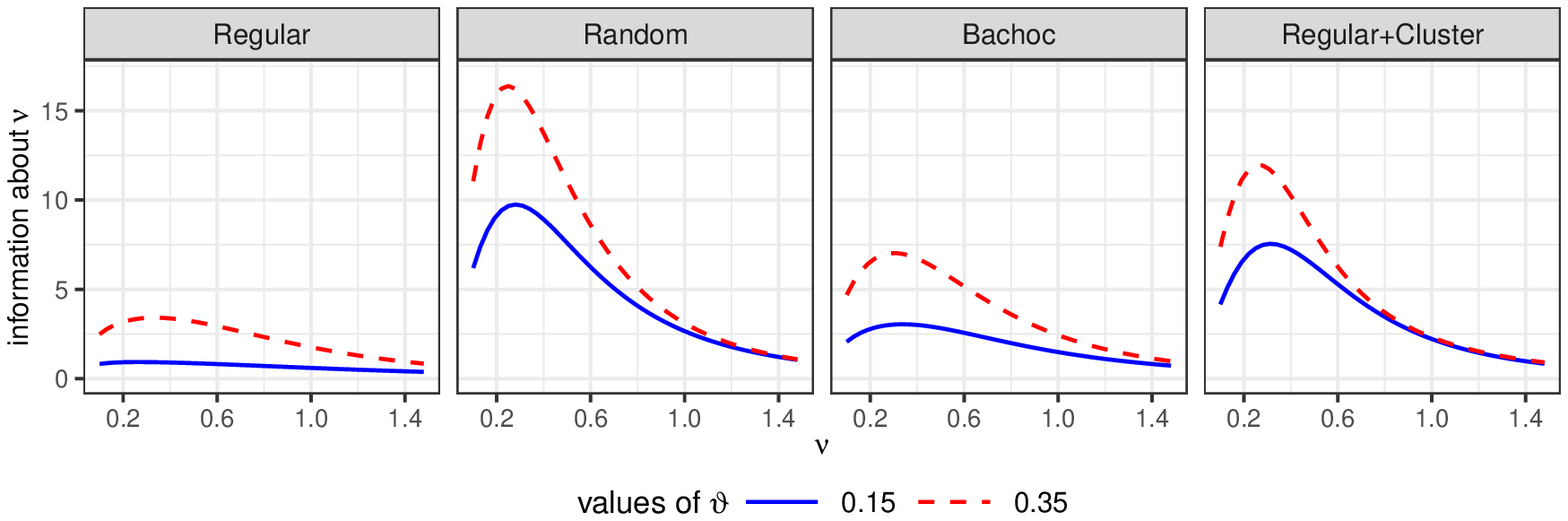, width=6in, height=2.5in}\\
\end{center}
\caption{Plots of the information about the smoothness parameter $\nu$, ${\rm Inf}(\mbox{\boldmath $\eta$},\mathcal{S}_n)_{4}$,
%$1/I(\mbox{\boldmath $\eta$},\mathcal{S}_n)^{\nu \nu}$,	
for different sampling designs, $\sigma^2 = 1$ and $\tau^2 = 0.2$. 
Top: information about $\nu$ as a function of $\vartheta$ for two values of $\nu$.	 
Bottom: information about $\nu$ as a function of $\nu$ for two values of $\vartheta$.} 
\label{fig:info-nu}
\end{figure}

\medskip

{\bf Information About the Smoothness Parameter}.
Next we investigate the variation of the information about the smoothness parameter $\nu$,
both as a function of  $\vartheta$ and as a function of $\nu$.

Figure \ref{fig:info-nu} (top panels) displays plots of 
${\rm Inf}(\mbox{\boldmath $\eta$},\mathcal{S}_n)_{4}$ 
%$1/I(\mbox{\boldmath $\eta$},\mathcal{S}_n)^{\nu \nu}$
as a function of $\vartheta$ when $\nu = 0.5$ and $1.5$ for the four sampling designs.
For both values of the smoothness parameter and all the designs, the information about $\nu$  
appears to increase monotonically when $\vartheta$ increases, and 
this information is larger when $\nu = 0.5$ than when $\nu = 1.5$.
When $\nu = 0.5$ the rate of increase in information is fast for small values of $\vartheta$, but 
slow for large values, to the point that the information becomes close to constant, especially for 
the Random and Regular+Cluster designs.
When $\nu = 1.5$ the information about $\nu$ is about constant and is small for all range parameters. 
Everything else being equal, the Regular design is the least informative about $\nu$ while 
the Random design is the most informative.
The presence of many nearby pairs of sampling locations in the latter design
enables better inference of smoothness.

Figure \ref{fig:info-nu} (bottom panels) displays plots of ${\rm Inf}(\mbox{\boldmath $\eta$},\mathcal{S}_n)_{4}$ 
%$1/I(\mbox{\boldmath $\eta$},\mathcal{S}_n)^{\nu \nu}$
as a function of $\nu$ when $\vartheta = 0.15$ and $0.35$ for the four sampling designs.
For both values of the range parameter and all designs, 
the information about $\nu$ generally increases up to a point and then decreases 
monotonically. Curiously, for all designs the information about $\nu$ peaks 
around the same value of $\nu$ ($\approx 0.25$).
In addition, regardless of the design and the smoothness parameter, the information about $\nu$
is larger when $\vartheta=0.35$ than when $\vartheta=0.15$.
Overall, this information becomes very small when the process is smooth, but for non--smooth processes
the information depends substantially on the design. 
Again, the Regular design is the least informative about $\nu$ while the Random design is 
the most informative.

\medskip

Figures \ref{fig:info-rho} and \ref{fig:info-nu} display the patterns of variation of 
the information about $\vartheta$ and $\nu$, as functions of $\vartheta$ or $\nu$, 
when $\sigma^2 = 1$ and $\tau^2 = 0.2$ are kept fixed.
Extensive numerical explorations (not shown) reveal that the same general patterns hold for 
other values of $\sigma^2$ and $\tau^2$.
Also, the same patterns of variation were observed for other sample sizes
($n = 100$ and $529$, not shown).
In addition, it was seen that the same patterns of variation displayed in 
Figures \ref{fig:info-rho} and \ref{fig:info-nu} hold when $\vartheta$ and $\nu$ are fixed at
other pairs of values; see also Figure \ref{fig:info-rho-nu} below.
These numerical explorations also suggest that, when everything else (model and design) is kept fixed, 
the information about $\vartheta$ is generally an increasing function of $\sigma^2$ and 
a decreasing function of $\tau^2$.
The information about $\nu$ also appears to increase with $\sigma^2$ and 
decrease with $\tau^2$, when everything else is kept fixed.

\begin{figure}[t!]
\begin{center}
\psfig{figure=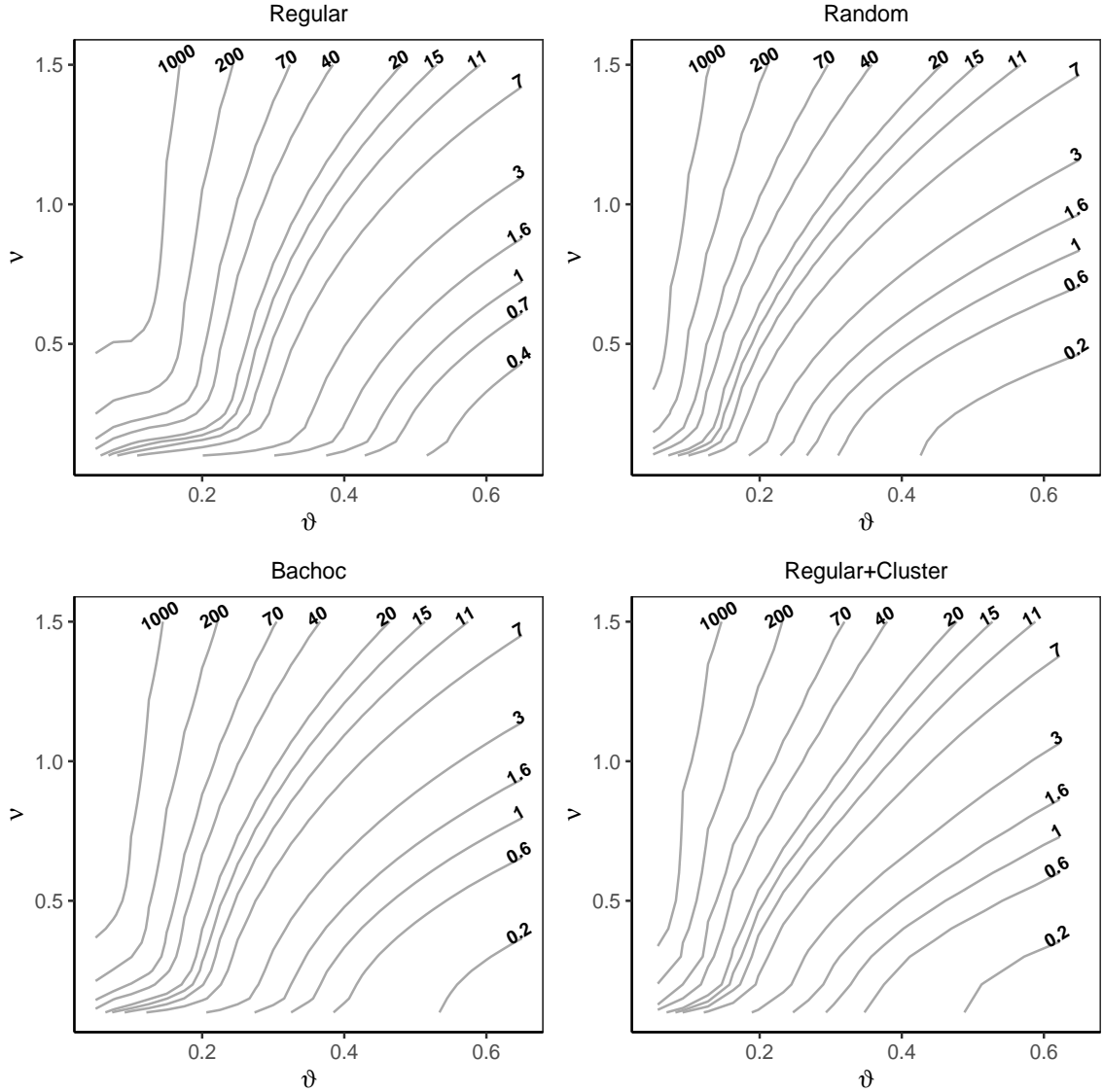, width=6in, height=6in}  
\end{center}
\caption{Contour plots of the ratio 
${\rm Inf}(\mbox{\boldmath $\eta$},\mathcal{S}_n)_{3}/
{\rm Inf}(\mbox{\boldmath $\eta$},\mathcal{S}_n)_{4}$
%$I(\mbox{\boldmath $\eta$},\mathcal{S}_n)^{\nu \nu}/I(\mbox{\boldmath $\eta$},\mathcal{S}_n)^{\vartheta \vartheta}$
for $\sigma^2 = 1$, $\tau^2 = 0.2$, $(\vartheta,\nu)$ in a grid of points in 
$[0.1, 0.9] \times [0.1, 1.5]$ and different sampling designs.} 
 \label{fig:info-rho-nu}
\end{figure}

\medskip

{\bf Information About the Range Parameter Relative to that of the Smoothness Parameter}.
Now we investigate the variation of the information about the range parameter relative to the information
about the smoothness parameter when $(\vartheta,\nu)$ varies over the 
region $[0.05, 0.65] \times [0.1, 1.5]$.
Figure \ref{fig:info-rho-nu} (top left) displays the contour plot of the ratio 
%of the third and fourth components of ${\rm Inf}(\mbox{\boldmath $\eta$},\mathcal{S}_n)$, 
${\rm Inf}(\mbox{\boldmath $\eta$},\mathcal{S}_n)_{3}/
{\rm Inf}(\mbox{\boldmath $\eta$},\mathcal{S}_n)_{4}$
%$I(\mbox{\boldmath $\eta$},\mathcal{S}_n)^{\nu \nu}/I(\mbox{\boldmath $\eta$},\mathcal{S}_n)^{\vartheta \vartheta}$,
when $\mathcal{S}_n$ is the regular design.
It shows that this ratio is less than 1 in a region of the correlation parameter space that combines a
large range parameter (strongly dependent process) and a small smoothness parameter (non--smooth process);
it can be loosely described as the `south--east' section of $[0.05, 0.65] \times [0.1, 1.5]$ 
and is denoted by $T_{\rm regular}$.
For these covariance models, the data contain substantial information about the smoothness parameter 
(when the information about the range parameter is used as reference).
%the two informations are not strictly comparable though}. 
On the other hand, the opposite holds in $T_{\rm regular}^c$, so for covariance models 
in this region the information about the smoothness parameter is less substantial.
The same overall behavior occurs for the Random, Bachoc and Regular+Cluster designs,
as shown in Figure \ref{fig:info-rho-nu}, with the ratio 
${\rm Inf}(\mbox{\boldmath $\eta$},\mathcal{S}_n)_{3}/
{\rm Inf}(\mbox{\boldmath $\eta$},\mathcal{S}_n)_{4}$
taking for each model values slightly smaller than those in the Regular design.
In addition, if $T_{\rm random}$,  $T_{\rm bachoc}$ and $T_{\rm reg+clu}$ are defined similarly as $T_{\rm regular}$,
it generally holds that $T_{\rm regular} \subset T_{\rm random}$, $T_{\rm regular} \subset T_{\rm bachoc}$ and 
$T_{\rm regular} \subset T_{\rm reg+clu}$.
Hence, for all the designs, the amount of information about the smoothness parameter is 
substantial when the true process is strongly dependent and non--smooth.
Once again, the same behaviors were observed when $\sigma^2$ and $\tau^2$ were fixed at other values.

\medskip
{\bf Information About Another Microergodic Parameter}.
As indicated by a referee, a parameter that may be considered as relevant as $\nu$ for spatial 
interpolation/prediction is $\zeta := \sigma^2/\vartheta^{2\nu}$, since it is also microergodic
\citep{Zhang2004}.
So we also investigate the pattern of variation of its information, both as a function of $\vartheta$ 
and as a function of $\nu$.
As for the other parameters, the information about $\zeta \; (= \zeta(\bfeta))$ 
is measured by the inverse of the Cramer--Rao lower bound for the variance of its unbiased estimators, 
which is given by \citep[page 76]{Keener2010}
\[
i(\bfeta, \mathcal{S}_n) := 
\big( \big(\nabla \zeta(\bfeta) \big)^\top I(\bfeta, \mathcal{S}_n \big)^{-1} \nabla \zeta(\bfeta) \big)^{-1} ,
\]
where
\[
\nabla \zeta(\bfeta) = \frac{1}{\vartheta^{2\nu}} \Bigg( 1, 0 , -\frac{2\sigma^2 \nu}{\vartheta}, 
-2\sigma^2 \log(\vartheta) \Bigg)^\top .
\]

\begin{figure}[t!]
\begin{center}
\psfig{figure=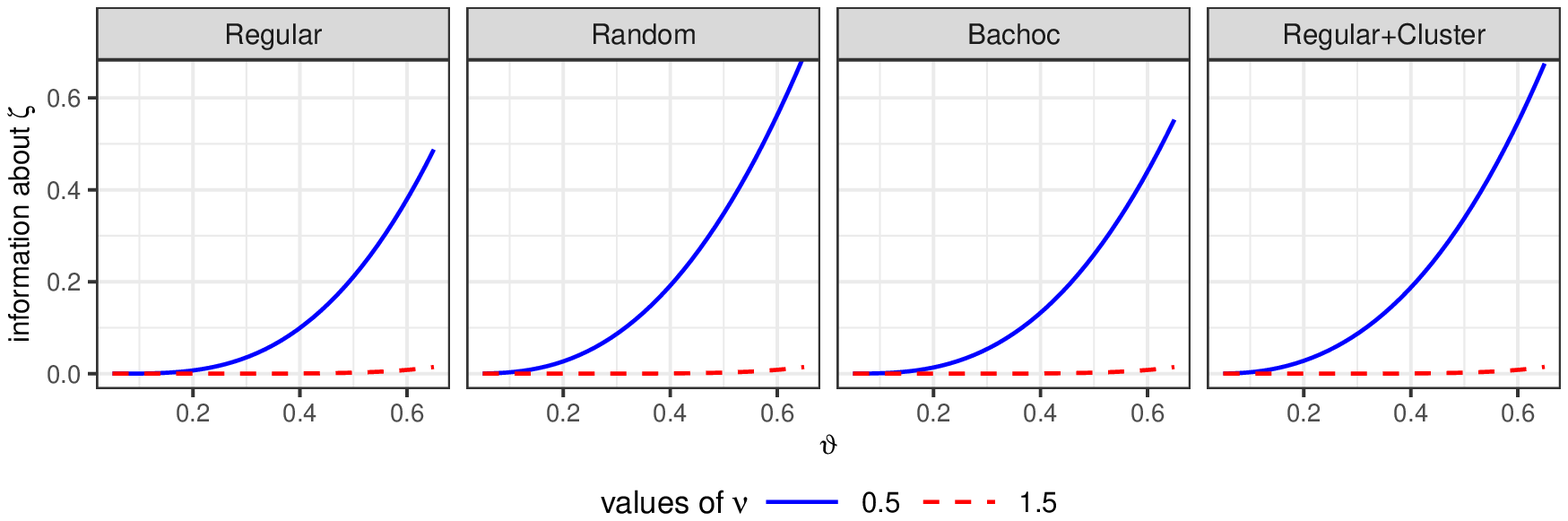, width=6in, height=2.5in}
\vspace{0.5 cm}
\psfig{figure=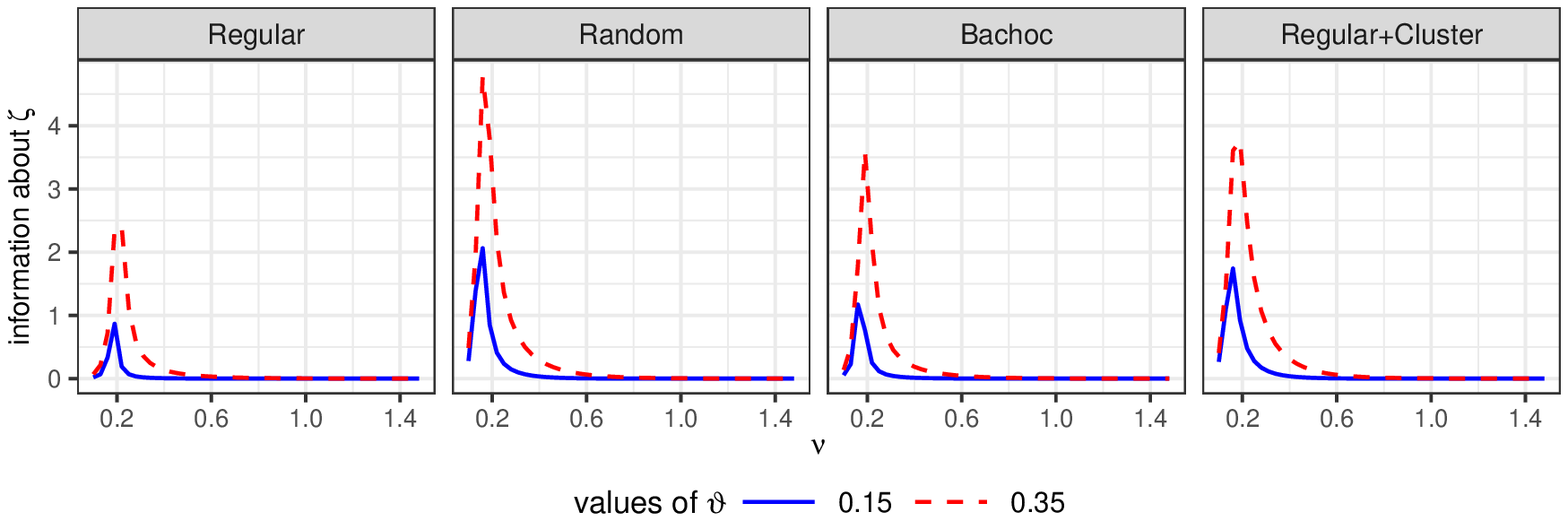, width=6in, height=2.5in}\\
\end{center}
\caption{
Plots of the information about the microergodic parameter $\zeta$, $i(\bfeta, \mathcal{S}_n)$,
for different sampling designs, $\sigma^2 = 1$ and $\tau^2 = 0.2$. 
Top: information about $\zeta$ as a function of $\vartheta$ for two values of $\nu$. 
Bottom: information about $\zeta$ as a function of $\nu$ for two values of $\vartheta$.} 
\label{fig:info-xi}
\end{figure}

Figure \ref{fig:info-xi} (top panels) displays plots of $i(\bfeta, \mathcal{S}_n)$ as a function of 
$\vartheta$ when $\nu = 0.5$ and $1.5$ for the four sampling designs.
For both values of the smoothness parameter and all the designs the information about $\zeta$ 
increases monotonically when $\vartheta$ increases.
Also, regardless of the design and the range parameter, the information about $\zeta$ is 
larger when $\nu=0.5$ than when $\nu=1.5$.
This information is not very sensitive to the design, although it is a bit larger for
the Random and Random+Cluster designs.

Figure \ref{fig:info-xi} (bottom panels) displays plots of $i(\bfeta, \mathcal{S}_n)$ as a function of $\nu$ 
when $\vartheta = 0.15$ and $0.35$ for the four sampling designs.
For both values of the range parameter and all the designs, the information about $\zeta$ 
increases up to a point and then decreases monotonically when $\nu$ increases. 
In addition, regardless of the design and the smoothness parameter, the information about $\zeta$ 
is larger when $\vartheta$ = 0.35 than when $\vartheta$ = 0.15.
Finally, the comparison of Figure \ref{fig:info-xi}  with 
Figure \ref{fig:info-nu} shows that the information about $\nu$ is larger than the information about $\zeta$ 
for all models and designs.

\medskip

\begin{figure}[t!]
\begin{center}
    \begin{tabular}{cc}
    \psfig{figure=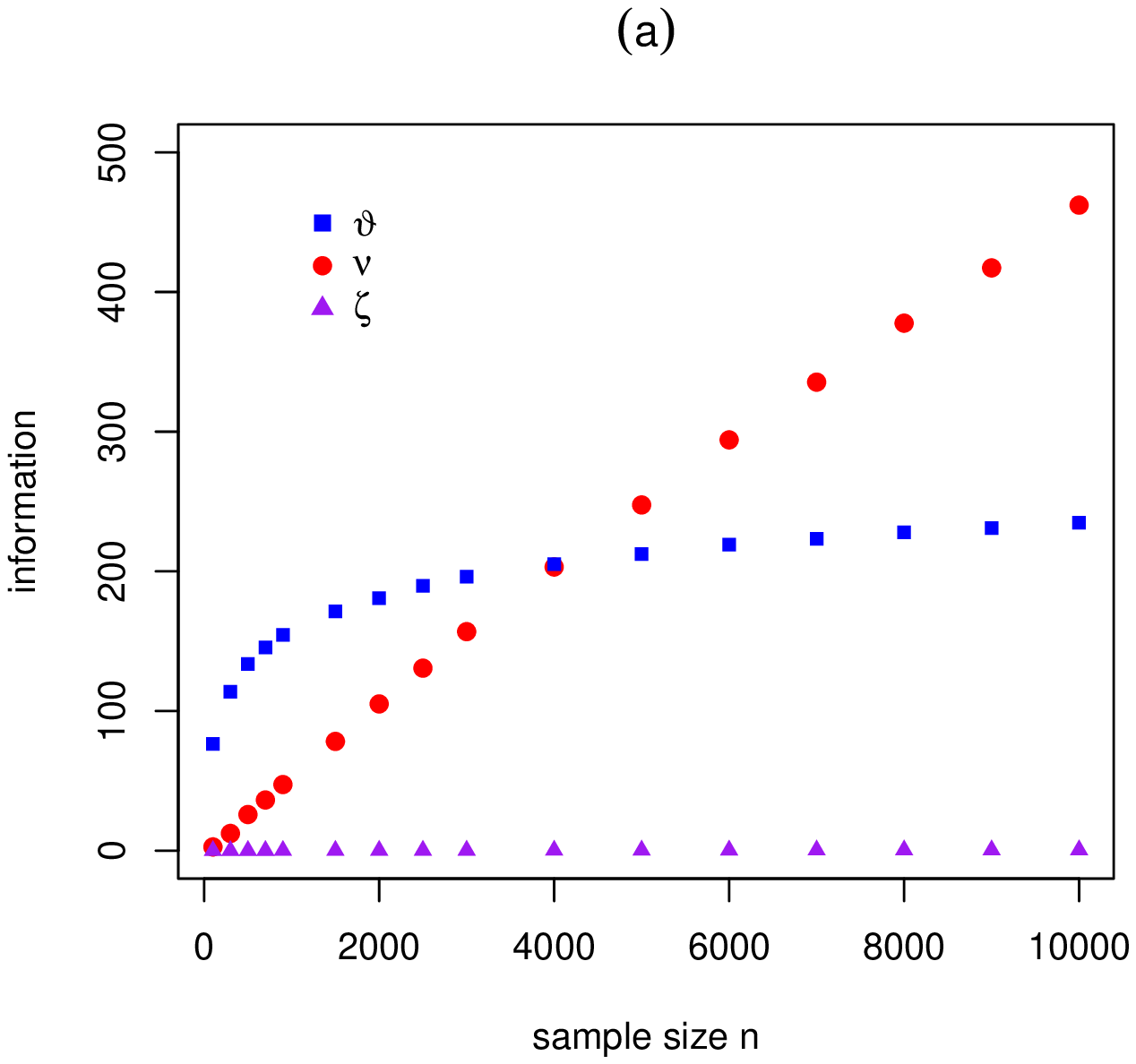, width=3in, height=3in} & 
    \psfig{figure=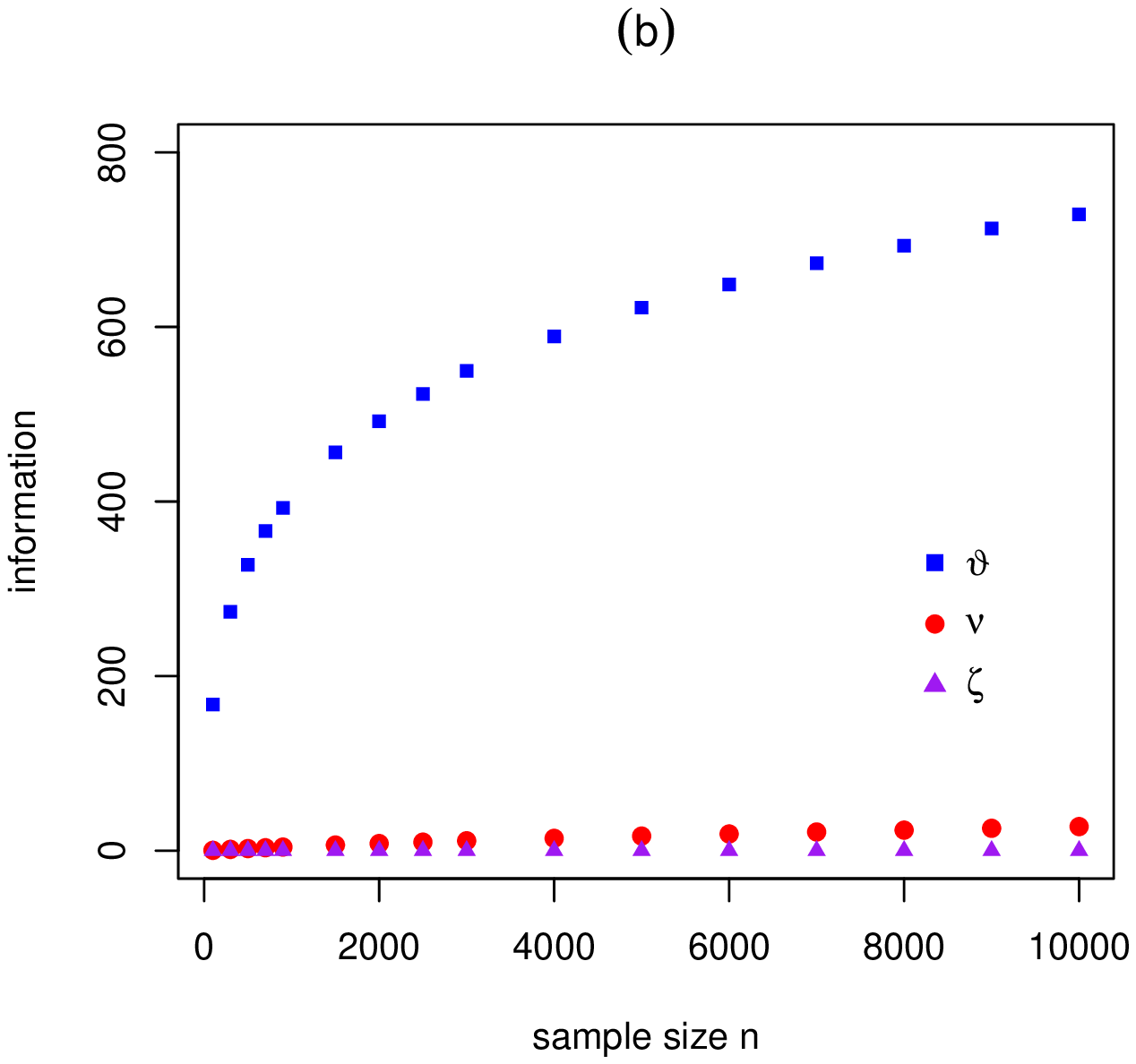, width=3in, height=3in}  
\\
    \psfig{figure=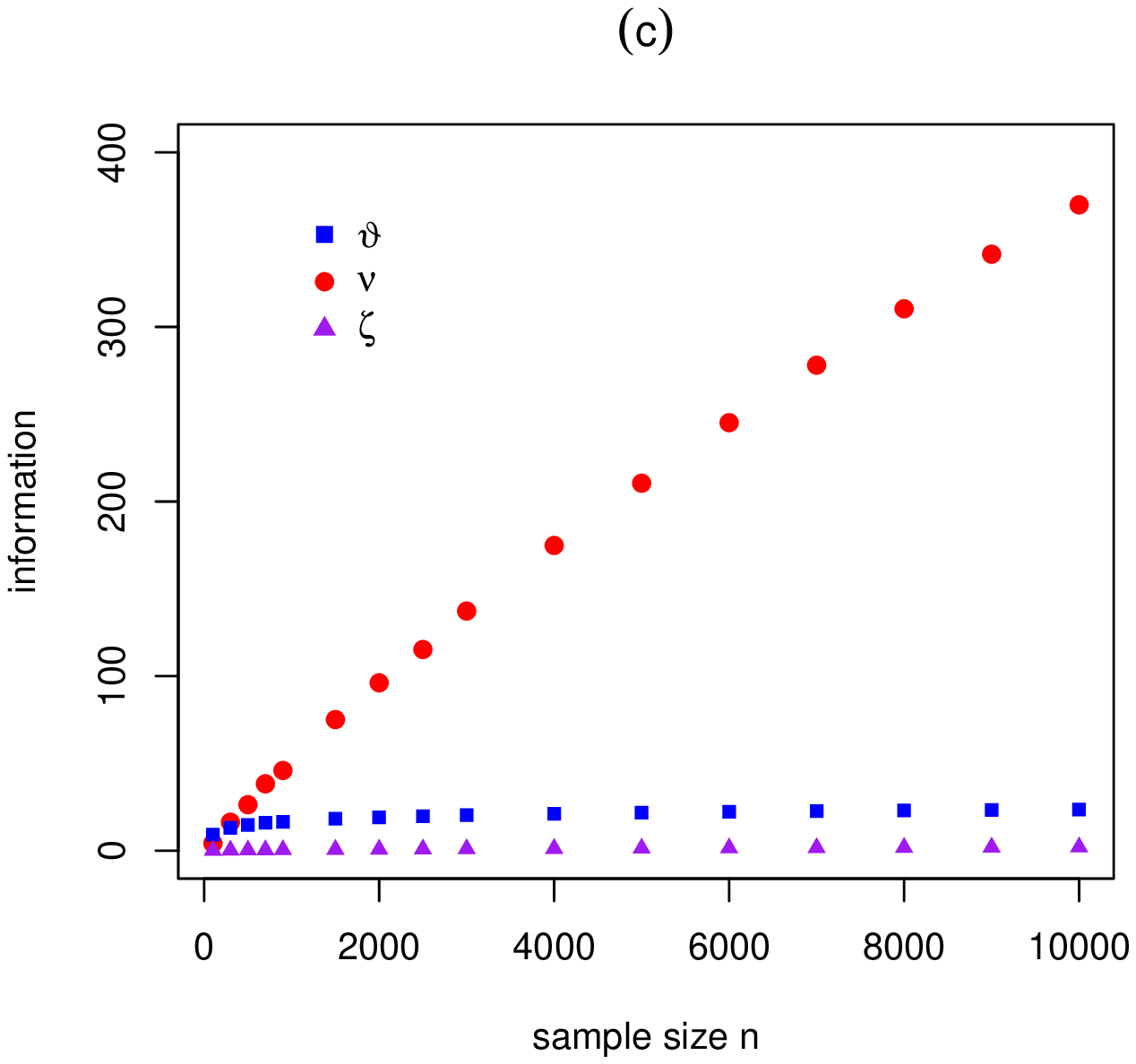, width=3in, height=3in} & 
    \psfig{figure=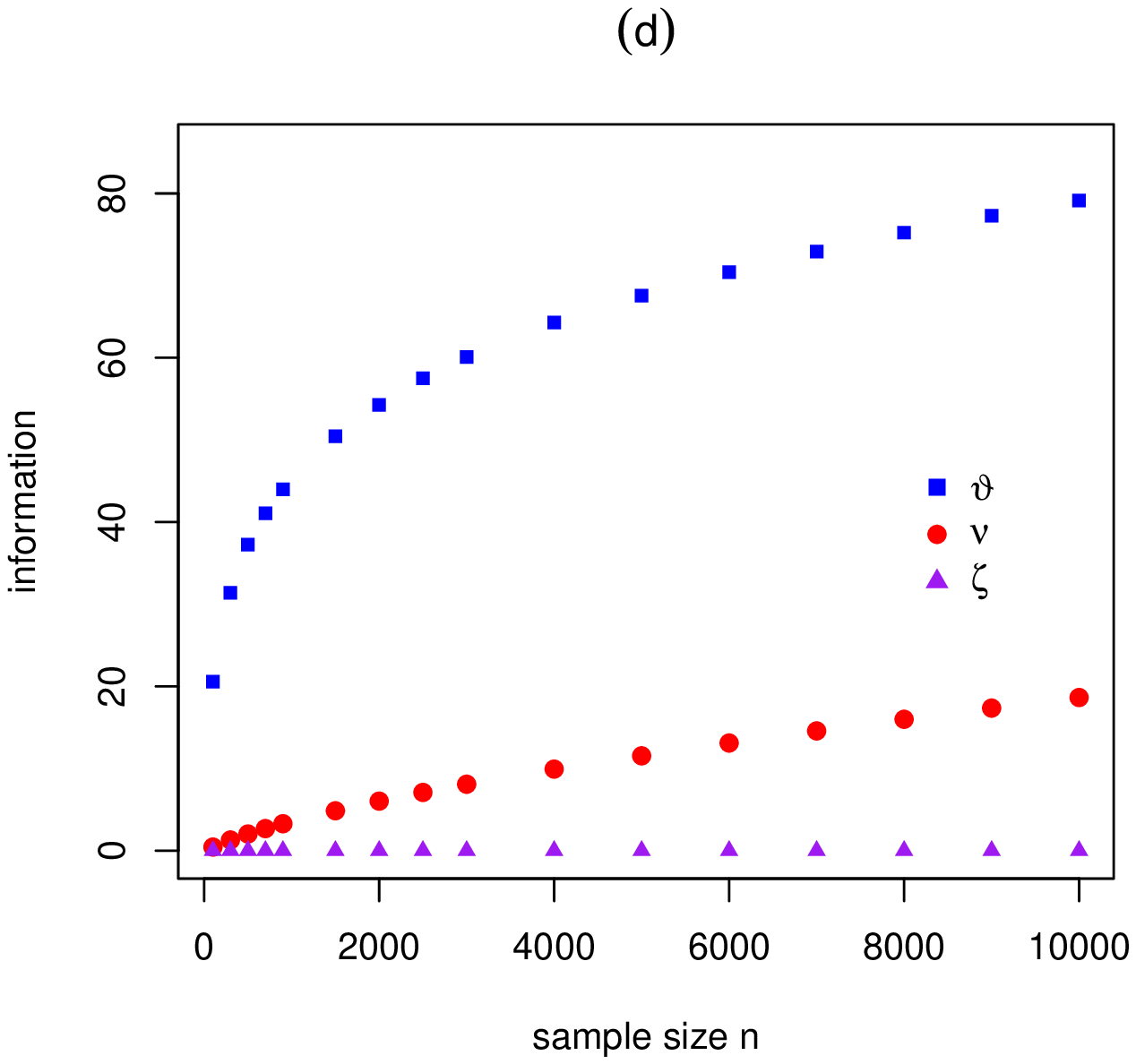, width=3in, height=3in}  
    \end{tabular} 
\end{center}
\caption{Plots of the information about $\vartheta$, $\nu$ and $\zeta$ versus sample size 
for the Random sampling design of size $n$ and the models: 
(a) $\bfeta = (1, 0.2, 0.2, 0.5)$, (b)  $\bfeta = (1, 0.2, 0.2, 1.5)$, 
(c) $\bfeta = (1, 0.2, 0.4, 0.5)$ and (d) $\bfeta = (1, 0.2, 0.4, 1.5)$.} 
\label{fig:info-th-nu-xi}
\end{figure}

{\bf Variation of Information With Sample Size}.
Figure \ref{fig:info-th-nu-xi} displays the variation of the information about $\vartheta$, $\nu$ and $\zeta$
with sample size for Random designs of size $n \leq 10000$ and the four models: 
(a) $\bfeta = (1, 0.2, 0.2, 0.5)$, (b)  $\bfeta = (1, 0.2, 0.2, 1.5)$, 
(c) $\bfeta = (1, 0.2, 0.4, 0.5)$ and (d) $\bfeta = (1, 0.2, 0.4, 1.5)$.
As expected, the information always increases with sample size, but the rate of change varies substantially
depending on the model. 
For instance, for models (a) and (c) [non--smooth models] the rate of change of the information about $\nu$ 
is larger than that of the information about $\vartheta$, while the opposite holds for models (b) and (d)
[smooth models], at least when $n \leq 10000$. 
These plots again show that for some models and sample sizes the information about $\nu$ is larger than 
that about $\vartheta$.
The information about $\zeta$ and its rate of change appear small for all models, showing again that
this information is smaller than the information about $\nu$.
The aforementioned behaviors were found to also hold for Regular designs of size $n \leq 10000$
(not shown).

\section{Local Model Influence}

An indirect but related approach to assess the information the data have about the covariance parameters 
is to quantify how sensitive the likelihood function is to small changes of the covariance parameters.
Assume $p=1$ and $\beta=0$, so $\bfeta$ are the only model parameters.
\cite{McCulloch1989} suggested a way to do this by carrying out an eigen--analysis of the (estimated) 
Fisher information matrix, which is related to the Kullback--Leibler divergence of the true model from
the estimated model. 
Specifically, let $(\lambda^{*}, \bfdelta^{*})$ be an eigenpair of 
$I(\mbox{\boldmath $\hat{\eta}$},\mathcal{S}_n)$, where $\lambda^*$ is its largest eigenvalue and
$\bfdelta^{*}$ is the corresponding eigenvector of unit Euclidean length.
Then, values of $\lambda^{*}$ larger than 1 indicate that the likelihood function is sensitive
to small changes of the covariance parameters, while the opposite holds when $\lambda^{*}$ is close to 0.
Additionally, the components of $\bfdelta^{*}$ indicate the coefficients of the linear combination of 
$\bfeta$ that is most influential.
See \cite{McCulloch1989} for details, and how to use this idea to assess the effects of small changes of
a prior distribution on the corresponding posterior and predictive distributions.

\subsection{A Telling Example (Continuation)}  \label{sec:telling-example-continuation}

We now revise and extend the analysis of the Swiss rainfall data in Section \ref{sec:telling-example}
using the exact Fisher information matrix computed as described in Section \ref{sec:fisher-info},
and re--scaling the spatial coordinates as described in 
Section \ref{sec:information-numerical-exploration}, with $r_{\rm max} = 335.71$ kilometers.
The inverse of the Fisher information matrix evaluated at the MLE 
$\mbox{\boldmath $\hat{\eta}$} = 
(\hat{\sigma}^2, \hat{\tau}^2, \hat{\vartheta}, \hat{\nu}) = (105.40, 6.72, 0.22, 0.96)$ is
\[
I(\mbox{\boldmath $\hat{\eta}$},\mathcal{S}_n)^{-1} = \left(\begin{array}{cccc}
1486.721 & -5.233 & 2.156 & -2.313 \\
-5.233 &  1.273 & -0.031 & 0.209 \\
2.156 & -0.031 & 0.005 & -0.013 \\ 
-2.313  & 0.209  & -0.013 & 0.064
\end{array}\right) .
\]
%\[(I(\mbox{\boldmath $\hat{\eta}$},\mathcal{S}_n))^{-1} = \left(\begin{array}{cccc}
%539.227 & -8.391 & 284.753 & -1.645 \\
%-8.391 & 4.279 & -13.424 & 0.450 \\
%284.753 & -13.424 & 265.309 & -3.110 \\
%-1.645 & 0.450 & -3.110 & 0.073
%\end{array}\right) ,
%\]
%which is not very close to the estimate in (\ref{rain-obs-info-inv}) provided by 
%the optimization algorithm. 
The (estimated) information about the covariance parameters is
\[
{\rm Inf}(\mbox{\boldmath $\hat{\eta}$},\mathcal{S}_n) =
(0.001, 0.786, 199.297, 15.643),
\]
%\[ {\rm Inf}(\mbox{\boldmath $\hat{\eta}$},\mathcal{S}_n) =
%(0.002,0.234,0.004,13.724)
%\]
confirming again that these data have substantial information about the smoothness parameter.
%than about the other covariance parameters.
The effect of re--scaling the spatial coordinates on the Fisher information matrix
is to produce large changes in the third row and third column while leaving the remaining entries
almost unchanged (when compared to not re--scaling).
This largely changes the information about the range parameter while leaving the information about
the other covariance parameters almost unaffected.

We also investigate the local influence of small changes of the covariance parameters on the 
likelihood. For the Swiss rainfall data the largest eigenvalue of 
$I(\mbox{\boldmath $\hat{\eta}$},\mathcal{S}_n)$ and its corresponding eigenvector are,
respectively,
\[
\lambda^{*} = 3.37 \times 10^3 \quad {\rm and} \quad \bfdelta^{*} = (-0.001, -0.014, 0.980, 0.200)^\top .
\]
This indicates that, for these data, the likelihood is sensitive to small changes of the
covariance parameters, since $\lambda^{*} >> 1$, and the range and smoothness parameters are the most influential,
since their corresponding coefficients in $\bfdelta^{*}$ have the largest magnitudes.

\section{Conclusions and Discussion}  \label{conclusions-discussion}

This work explored the patterns of variation of the information about the correlation parameters
of the Mat\'ern model in Gaussian random fields. 
Using the inverse diagonal elements of the inverse of the Fisher information matrix as 
an information measure, it was shown that the information about these correlation parameters varies 
substantially depending on the true model and sampling design.
Specifically, the following conclusions follow from the numerical explorations in 
Section \ref{sec:information-numerical-exploration}:

\begin{enumerate}
\item
In general, the information about the range parameter $\vartheta$ displays little sensitivity to
the sampling design.

\item
Except for weakly correlated processes observed on a regular design, the information about $\vartheta$ 
decreases when $\vartheta$ increases, to the point of being very small for processes with strong correlation.

\item
The information about $\vartheta$ increases when $\nu$ increases. 
This information is largest for processes with weak correlation that are smooth.

\item
The information about the smoothness parameter $\nu$ does display sensitivity to the sampling design.
The Regular design is the least informative about $\nu$, while the Random design is the most informative.

\item
The information about $\nu$ increases when $\vartheta$ increases.
This information is the largest for processes with strong correlation that are non--smooth.

\item
The information about $\nu$ does not display a monotonic pattern of change with $\nu$.

\item
Overall and regardless of the design, the information about the smoothness parameter $\nu$ is substantial for processes with 
strong correlation that are non--smooth.

\item
The information about the microergodic parameter $\zeta$ increases 
when $\vartheta$ increases and does not changes monotonically when $\nu$ increases. 
In general this information is smaller than the information about $\nu$.

\item
The information about $\vartheta$, $\nu$ and $\zeta$ grows with sample size, but
the rate of change varies substantially with the true model. 
The rate of change of information about $\nu$ is fastest for non--smooth processes, while the
rate of change of information about $\zeta$ appear always small. 

\end{enumerate}

Some of the above conclusions confirm  similar empirical findings reported in the literature, 
while other conclusions 
cast doubts on the unqualified claim regarding
geostatistical data having little or no information about the smoothness parameter of the
Mat\'ern model, as well as the common practice of fixing this parameter at some customary value
(e.g., $\nu = 0.5$ or $1.5$).
The analysis of the Swiss rainfall data provided an example where the data contain substantial 
information about the smoothness parameter.
These findings call for a reassessment of this practice and a shift toward a more 
statistically sound practice where inferences about the random field rely on data--based estimates 
of both correlation parameters, at least for strongly dependent and non--smooth processes observed 
on irregular sampling designs. 
For instance, \cite{McCullagh-Clifford2006} provided evidence showing that data
from crop yield processes tend to possess these features.

Although the above findings were obtained for Gaussian random fields, they also hold for
some non--Gaussian random fields. 
For instance, this is the case for some transformed Gaussian random fields (e.g., log--Gaussian random fields)
since the Fisher information matrix is invariant under differentiable one--to--one transformations
of the data.
Also, although the study was restricted to the Mat\'ern family of covariance functions, 
we conjecture that the above findings also hold for other families of covariance functions, 
such as the generalized Wendland \citep{Bevilacqua2019} and power exponential \citep{DeOliveira1997} 
families that, as the Mat\'ern, depend on a smoothness or roughness parameter.

The approach presented in Section \ref{sec:fisher-info} to compute 
$({\partial}/{\partial \nu}) K_{\boldsymbol{\vartheta}}(r)$ might not be fast or accurate enough
to compute the expected or observed Fisher information matrices used in iterative algorithms 
for the computation of maximum likelihood estimates based on large data sets.
Recently, \cite{Geoga-etal-2022} developed methods and software for fast and accurate 
computation of this derivative using automatic differentiation based on a new implementation of 
the Bessel function $\mathcal{K}_{\nu}(x)$.
These should contribute to ameliorate the computational challenges of estimating all the parameters
in the Mat\'ern family.
\bigskip

\section*{Acknowledgment}

The authors thank the Associate Editor and three  anonymous reviewers for their insightful comments
and suggestions that lead to an improved article. We also thank Eric Slud for stimulating
conversations and feedback in the early stages of this research. Victor De Oliveira was partially
supported by the U.S. National Science Foundation grant DMS--2113375. Zifei Han was supported by the Fundamental Research Funds for the Central Universities, China in UIBE (CXTD11-05). 
\bigskip

\section*{Appendix}  \label{sec:app}

\noindent
{\bf Derivation of Identities (\ref{deriv-corr-mat-vartheta}) and (\ref{deriv-corr-mat-nu})}

To derive (\ref{deriv-corr-mat-vartheta}), write the Mat\'ern correlation function as 
$K_{\boldsymbol{\bfvartheta}}(r) = c(\nu) b(\vartheta)^{\nu} 
\mathcal{K}_{\nu}(b(\vartheta))$, where 
$c(\nu) := {2^{1 - \nu}}/{\Gamma(\nu)}$ and $b(\vartheta) := {2r\sqrt{\nu}}/{\vartheta} $.
Then by direct differentiation
\begin{align*}
\frac{\partial}{\partial \vartheta} K_{\boldsymbol{\vartheta}}(r) &=
c(\nu) \Big( \nu b(\vartheta)^{\nu - 1} b'(\vartheta) \mathcal{K}_{\nu}(b(\vartheta))
+  b(\vartheta)^{\nu} \frac{\partial}{\partial x} \mathcal{K}_{\nu}(x) \Big|_{x=b(\vartheta)}
\!\! \cdot b'(\vartheta) \Big) \\
&= c(\nu) b(\vartheta)^{\nu - 1} b'(\vartheta) \Big( \nu \mathcal{K}_{\nu}(b(\vartheta))
	-b(\vartheta)\Big[ \mathcal{K}_{\nu - 1}(b(\vartheta)) +
	\frac{\nu \vartheta}{2\sqrt{\nu} r} \mathcal{K}_{\nu}(b(\vartheta))  \Big] \Big) \\
	&= -\frac{c(\nu) (2\sqrt{\nu}r)^{\nu}}{\vartheta^{\nu + 1}} \Bigg(
\nu \mathcal{K}_{\nu}\Big( \frac{2\sqrt{\nu}}{\vartheta} r \Big) 
-\frac{2\sqrt{\nu}}{\vartheta} r \mathcal{K}_{\nu - 1}\Big( \frac{2\sqrt{\nu}}{\vartheta} r \Big)
-\nu \mathcal{K}_{\nu}\Big( \frac{2\sqrt{\nu}}{\vartheta} r \Big) \Bigg) \\
	&= \frac{4 \nu^{\frac{\nu + 1}{2}} r^{\nu+1}}{\Gamma(\nu)\vartheta^{\nu + 2}} 
\mathcal{K}_{\nu - 1}\Big( \frac{2\sqrt{\nu}}{\vartheta} r \Big)  ,
\end{align*}
where the second identity follows from (\ref{deriv-K-x}).

To derive (\ref{deriv-corr-mat-nu}), write the Mat\'ern correlation function as
\[
K_{\boldsymbol{\vartheta}}(r) = 2 e^{\nu \log( 1/2)} (\Gamma(\nu))^{-1}
e^{\nu \log( \frac{2r}{\vartheta} \sqrt{\nu})}
\mathcal{K}_{\nu}\Big( \frac{2r}{\vartheta} \sqrt{\nu} \Big) ,
\] 
so after direct differentiation we have
\begin{equation}
	\frac{\partial}{\partial \nu} K_{\boldsymbol{\vartheta}}(r) =
\left(\frac{1}{2} + \log\left(\frac{\sqrt{\nu}}{\vartheta} r\right) - \psi(\nu)\right)
K_{\boldsymbol{\vartheta}}(r) 
+ h(\nu) \frac{\partial}{\partial \nu} \mathcal{K}_{\nu}\Big( \frac{2r}{\vartheta} \sqrt{\nu} \Big) .
\label{deriv-corr-mat-nu1}
\end{equation}
Now, let $G(x, y) :=  \mathcal{K}_{x}(y)$. Then
\begin{align}
	\frac{\partial}{\partial \nu} \mathcal{K}_{\nu}\Big( \frac{2r}{\vartheta} \sqrt{\nu} \Big) &=
\frac{\partial}{\partial \nu} G\Big(\nu,  \frac{2r}{\vartheta} \sqrt{\nu}\Big) \nonumber \\
&= \frac{\partial}{\partial x} G(x, y) \Big|_{x=\nu, y=\frac{2r}{\vartheta} \sqrt{\nu}} +\;
\frac{\partial}{\partial y} G(x, y) \Big|_{x=\nu, y=\frac{2r}{\vartheta} \sqrt{\nu}} 
	\cdot \frac{r}{\vartheta \sqrt{\nu}}  \nonumber \\
	&= \int_{0}^{\infty} t \sinh(\nu t) e^{-\frac{2r}{\vartheta} \sqrt{\nu} \cosh(t)} dt \nonumber \\
& \quad\quad\quad\quad
	-\frac{r}{\vartheta \sqrt{\nu}}  \Bigg( \mathcal{K}_{\nu - 1}\Big( \frac{2r}{\vartheta} \sqrt{\nu} \Big)
+ \frac{\vartheta \sqrt{\nu}}{2r} \mathcal{K}_{\nu}\Big( \frac{2r}{\vartheta} \sqrt{\nu} \Big) \Bigg) 
\nonumber \\
&= \int_{0}^{\infty} t \sinh(\nu t) e^{-\frac{2r}{\vartheta} \sqrt{\nu} \cosh(t)} dt
-\frac{r}{\vartheta \sqrt{\nu}} \mathcal{K}_{\nu - 1}\Big( \frac{2r}{\vartheta} \sqrt{\nu} \Big)
-\frac{1}{2} \mathcal{K}_{\nu}\Big( \frac{2r}{\vartheta} \sqrt{\nu} \Big) ,
\label{deriv-corr-mat-nu2}
\end{align}
where the third identity follows from (\ref{deriv-K-x}) and (\ref{deriv-K-nu}).
Finally, replacing (\ref{deriv-corr-mat-nu2}) into (\ref{deriv-corr-mat-nu1}) we get
\begin{align*}
\frac{\partial}{\partial \nu} K_{\boldsymbol{\vartheta}}(r) &= \frac{1}{2}K_{\boldsymbol{\vartheta}}(r)
+ \left(\log\left(\frac{\sqrt{\nu}}{\vartheta} r\right) - \psi(\nu)\right) K_{\boldsymbol{\vartheta}}(r) 
+  h(\nu) \int_{0}^{\infty} t \sinh(\nu t) e^{-\frac{2r}{\vartheta} \sqrt{\nu} \cosh(t)} dt \\
& \quad\quad
- h(\nu) \frac{r}{\vartheta \sqrt{\nu}} \mathcal{K}_{\nu - 1}\Big( \frac{2r}{\vartheta} \sqrt{\nu} \Big)
- \frac{1}{2} 
\underbrace{h(\nu) \mathcal{K}_{\nu}\Big( \frac{2r}{\vartheta} \sqrt{\nu} \Big)}_{K_{\boldsymbol{\vartheta}}(r)} \\
&= \left(\log\left(\frac{\sqrt{\nu}}{\vartheta} r\right) - \psi(\nu)\right) K_{\boldsymbol{\vartheta}}(r) \\
& \quad\quad\quad 
- h(\nu) \Big(\frac{r}{\vartheta \sqrt{\nu}} \mathcal{K}_{\nu - 1}\Big( \frac{2r}{\vartheta} \sqrt{\nu} \Big) 
- \int_{0}^{\infty} t \sinh(\nu t) e^{-\frac{2r}{\vartheta} \sqrt{\nu} \cosh(t)} dt\Big) .
\end{align*}

\bigskip
\bibliography{JABES-Final.bib}  

\begin{thebibliography}{}

\bibitem[\protect\citeauthoryear{Bachoc}{Bachoc}{2014}]{Bachoc2014}
Bachoc, F. (2014).
\newblock Asymptotic analysis of the role of spatial sampling for covariance
  parameter estimation of {Gaussian} processes.
\newblock {\em Journal of Multivariate Analysis\/}~{\em 125}, 1--35.

\bibitem[\protect\citeauthoryear{Bevilacqua, Faouzi, Furrer, and
  Porcu}{Bevilacqua et~al.}{2019}]{Bevilacqua2019}
Bevilacqua, M., T.~Faouzi, R.~Furrer, and E.~Porcu (2019).
\newblock Estimation and prediction using generalized {Wendland} covariance
  functions under fixed domain asymptotics.
\newblock {\em The Annals of Statistics\/}~{\em 47}, 828--856.

\bibitem[\protect\citeauthoryear{Bose, Hodges, and Banerjee}{Bose
  et~al.}{2018}]{Bose2018}
Bose, M., J.~Hodges, and S.~Banerjee (2018).
\newblock Toward a diagnostic toolkit for linear models with
  {Gaussian}--process distributed random effects.
\newblock {\em Biometrics\/}~{\em 74}, 863--873.

\bibitem[\protect\citeauthoryear{Cressie}{Cressie}{1993}]{Cressie1993}
Cressie, N. (1993).
\newblock {\em Statistics for Spatial Data (rev. ed.)}.
\newblock Wiley.

\bibitem[\protect\citeauthoryear{De~Oliveira, Kedem, and Short}{De~Oliveira
  et~al.}{1997}]{DeOliveira1997}
De~Oliveira, V., B.~Kedem, and D.~Short (1997).
\newblock Bayesian prediction of transformed {Gaussian} random fields.
\newblock {\em Journal of the American Statistical Association\/}~{\em 92},
  1422--1433.

\bibitem[\protect\citeauthoryear{Diggle and Lophaven}{Diggle and
  Lophaven}{2006}]{Diggle-Lophaven2006}
Diggle, P. and S.~Lophaven (2006).
\newblock Bayesian geostatistical design.
\newblock {\em Scandinavian Journal of Statistics\/}~{\em 33}, 53--64.

\bibitem[\protect\citeauthoryear{Diggle and Ribeiro}{Diggle and
  Ribeiro}{2007}]{Diggle2007}
Diggle, P. and P.~Ribeiro (2007).
\newblock {\em Model--Based Geostatistics}.
\newblock Springer--Verlag.

\bibitem[\protect\citeauthoryear{Gelfand and Schliep}{Gelfand and
  Schliep}{2016}]{Gelfand2016}
Gelfand, A. and E.~Schliep (2016).
\newblock Spatial statistics and {Gaussian} processes: A beautiful marriage.
\newblock {\em Spatial Statistics\/}~{\em 18}, 86--104.

\bibitem[\protect\citeauthoryear{Geoga, Marin, Schanen, and Stein}{Geoga
  et~al.}{2022}]{Geoga-etal-2022}
Geoga, C., O.~Marin, M.~Schanen, and M.~Stein (2022).
\newblock Fitting {M}at\'ern smoothness parameters using automatic
  differentiation.
\newblock {\em arXiv:2201.00090v1\/}.

\bibitem[\protect\citeauthoryear{Gradshteyn and Ryzhik}{Gradshteyn and
  Ryzhik}{2000}]{Gradshteyn2000}
Gradshteyn, T. and I.~Ryzhik (2000).
\newblock {\em Table of Integrals, Series, and Products, 6th edition}.
\newblock Academic Press.

\bibitem[\protect\citeauthoryear{Guttorp and Gneiting}{Guttorp and
  Gneiting}{2006}]{Guttorp2006}
Guttorp, P. and T.~Gneiting (2006).
\newblock Studies in the history of probability and statistics {XLIX} on the
  {M}at\'ern correlation family.
\newblock {\em Biometrika\/}~{\em 93}, 989--995.

\bibitem[\protect\citeauthoryear{Handcock and Stein}{Handcock and
  Stein}{1993}]{Handcock1993}
Handcock, M. and M.~Stein (1993).
\newblock A {Bayesian} analysis of kriging.
\newblock {\em Technometrics\/}~{\em 35}, 403--410.

\bibitem[\protect\citeauthoryear{Kaufman and Shaby}{Kaufman and
  Shaby}{2013}]{Kaufman2013}
Kaufman, C. and B.~Shaby (2013).
\newblock The role of the range parameter for estimation and prediction in
  geostatistics.
\newblock {\em Biometrika\/}~{\em 100}, 473--484.

\bibitem[\protect\citeauthoryear{Keener}{Keener}{2010}]{Keener2010}
Keener, R. (2010).
\newblock {\em Theoretical Statistics: Topics for a Core Course}.
\newblock Springer.

\bibitem[\protect\citeauthoryear{Loh}{Loh}{2015}]{Loh2015}
Loh, W.-L. (2015).
\newblock Estimating the smoothness of a {G}aussian random field from
  irregularly spaced data via higher--order quadratic variations.
\newblock {\em The Annals of Statistics\/}~{\em 43}, 2766--2794.

\bibitem[\protect\citeauthoryear{Loh, Sun, and Wen}{Loh
  et~al.}{2021}]{Loh-etal-2021}
Loh, W.-L., S.~Sun, and J.~Wen (2021).
\newblock On fixed--domain asymptotics, parameter estimation and isotropic
  {G}aussian random fields with {M}at\'ern covariance functions.
\newblock {\em The Annals of Statistics\/}~{\em 49}, 3127--3152.

\bibitem[\protect\citeauthoryear{Mat\'ern}{Mat\'ern}{1986}]{Matern1986}
Mat\'ern, B. (1986).
\newblock {\em Spatial Variation\/} (2nd ed.).
\newblock Springer--Verlag.

\bibitem[\protect\citeauthoryear{McCullagh and Clifford}{McCullagh and
  Clifford}{2006}]{McCullagh-Clifford2006}
McCullagh, P. and D.~Clifford (2006).
\newblock Evidence for conformal invariance of crop yields.
\newblock {\em Proceedigns of the Royal Society A\/}~{\em 462}, 2119--2143.

\bibitem[\protect\citeauthoryear{McCulloch}{McCulloch}{1989}]{McCulloch1989}
McCulloch, R. (1989).
\newblock Local model influence.
\newblock {\em Journal of the American Statistical Association\/}~{\em 84},
  473--478.

\bibitem[\protect\citeauthoryear{Papritz and Schwierz}{Papritz and
  Schwierz}{2021}]{Papritz2021}
Papritz, A. and C.~Schwierz (2021).
\newblock {\tt georob}: Robust geostatistical analysis of spatial data.
\newblock {\em R package version 0.3-14\/}.

\bibitem[\protect\citeauthoryear{Seber and Wild}{Seber and
  Wild}{2003}]{Seber2003}
Seber, G. and C.~Wild (2003).
\newblock {\em Nonlinear Regression}.
\newblock Wiley.

\bibitem[\protect\citeauthoryear{Stein}{Stein}{1988}]{Stein1988}
Stein, M. (1988).
\newblock Asymptotically efficient prediction of a random field with a
  misspecified covariance function.
\newblock {\em The Annals of Statistics\/}~{\em 16}, 55--63.

\bibitem[\protect\citeauthoryear{Stein}{Stein}{1990}]{Stein1990}
Stein, M. (1990).
\newblock Uniform asymptotic optimality of linear predictions of a random field
  using an incorrect second--order structure.
\newblock {\em The Annals of Statistics\/}~{\em 18}, 850--872.

\bibitem[\protect\citeauthoryear{Stein}{Stein}{1993}]{Stein1993}
Stein, M. (1993).
\newblock A simple condition for asymptotic optimality of linear predictions of
  random fields.
\newblock {\em Statistics \& Probability Letters\/}~{\em 17}, 399--404.

\bibitem[\protect\citeauthoryear{Stein}{Stein}{1999}]{Stein1999}
Stein, M. (1999).
\newblock {\em Interpolation of Spatial Data: Some Theory for Kriging}.
\newblock Springer--Verlag.

\bibitem[\protect\citeauthoryear{Steiner, Houze, and Yuter}{Steiner
  et~al.}{1995}]{Steiner1995}
Steiner, M., R.~Houze, and S.~Yuter (1995).
\newblock Climatological characterization of three--dimensional structure from
  operational radar and gauge data.
\newblock {\em Journal of Applied Meteorology\/}~{\em 34}, 1978--2007.

\bibitem[\protect\citeauthoryear{Wu and Lim}{Wu and Lim}{2016}]{Wu2016}
Wu, W.-Y. and C.~Lim (2016).
\newblock Estimation of smoothness of a stationary {G}aussian random field.
\newblock {\em Statistica Sinica\/}~{\em 26}, 1729--1745.

\bibitem[\protect\citeauthoryear{Wu, Lim, and Xiao}{Wu et~al.}{2013}]{Wu2013}
Wu, W.-Y., C.~Lim, and Y.~Xiao (2013).
\newblock Tail estimation of the spectral density for a stationary {G}aussian
  random field.
\newblock {\em Journal of Multivariate Analysis\/}~{\em 116}, 74--91.

\bibitem[\protect\citeauthoryear{Zhang}{Zhang}{2004}]{Zhang2004}
Zhang, H. (2004).
\newblock Inconsistent estimation and asymptotically equal interpolations in
  model--based geostatistics.
\newblock {\em Journal of the American Statistical Association\/}~{\em 99},
  250--261.

\bibitem[\protect\citeauthoryear{Zhu and Stein}{Zhu and
  Stein}{2005}]{Zhu-Stein2005}
Zhu, Z. and M.~Stein (2005).
\newblock Spatial sampling design for parameter estimation of the covariance
  function.
\newblock {\em Journal of Statistical Planning and Inference\/}~{\em 134},
  583--603.

\bibitem[\protect\citeauthoryear{Zhu and Stein}{Zhu and
  Stein}{2006}]{Zhu-Stein2006}
Zhu, Z. and M.~Stein (2006).
\newblock Spatial sampling design for prediction with estimated parameters.
\newblock {\em Journal of Agricultural, Biological, and Environmental
  Statistics\/}~{\em 11}, 24--44.

\bibitem[\protect\citeauthoryear{Zhu and Zhang}{Zhu and
  Zhang}{2006}]{Zhu-Zhang2006}
Zhu, Z. and H.~Zhang (2006).
\newblock Spatial sampling design under the infill asymptotic framework.
\newblock {\em Environmetrics\/}~{\em 17}, 323--337.

\bibitem[\protect\citeauthoryear{Zimmerman}{Zimmerman}{2006}]{Zimmerman2006}
Zimmerman, D. (2006).
\newblock Optimal network design for spatial prediction, covariance parameter
  estimation, and empirical prediction.
\newblock {\em Environmetrics\/}~{\em 17}, 635--652.

\bibitem[\protect\citeauthoryear{Zimmerman}{Zimmerman}{2010}]{Zimmerman2010}
Zimmerman, D. (2010).
\newblock {\em Likelihood--Based Methods. In: Handbook of Spatial Statistics,
  A.E. Gelfand, P.J. Diggle, M. Fuentes and P. Guttorp (eds.)}.
\newblock CRC Press, 45-56.

\end{thebibliography}


\begin{thebibliography}{100}

%\bibitem{Abramowitz1964}
%Abramowitz, M. and Stegun, I.A. (1964),
%{\it Handbook of Mathematical Functions with Formulas, Graphs, and Mathematical Tables}.

\bibitem{Bevilacqua-etal2019}
Bevilacqua, M., Faouzi, T., Furrer, R. and Porcu, E. (2019), Estimation and Prediction Using Generalized
Wendland Covariance Functions Under Fixed Domain Asymptotics, {\it Annals of Statistics}, 47, 828-856.	

\bibitem{Bose2018}
Bose, M., Hodges, J.S. and Banerjee, S. (2018), Toward a Diagnostic Toolkit for Linear Models 
with Gaussian--Process Distributed Random Effects, {\it Biometrics}, 74, 863-873.

\bibitem{HoetingCooley2011}
Cooley, D. and Hoeting, J.A. (2011), Discussion of `An Explicit Link Between Gaussian Fields and
Gaussian Markov Random Fields: The Stochastic Partial Differential Equation Approach,'
{\it Journal of the Royal Statistical Society--Series B}, 73, 470.

\bibitem{Cressie1993}
Cressie, N. (1993), {\it Statistics for Spatial Data} (rev. ed.), Wiley. 

\bibitem{DeOliveira1997}
De Oliveira, V., Kedem, B. and Short, D.A. (1997), Bayesian Prediction of Transformed 
Gaussian Random Fields, 
{\it Journal of the American Statistical Association}, 92, 1422-1433.

\bibitem{Diggle2007}
Diggle, P.J. and Ribeiro, P.J. (2007), {\it Model--Based Geostatistics}. Springer--Verlag.

\bibitem{Diggle-Lophaven2006}
Diggle, P. and Lophaven, S. (2006), Bayesian Geostatistical Design,
{\it Scandinavian Journal of Statistics}, 33, 53-64.

\bibitem{Gelfand2016}
Gelfand, A.E. and Schliep, E.M. (2016), Spatial Statistics and Gaussian Processes:
A Beautiful Marriage, {\it Spatial Statistics}, 18, 86-104.

\bibitem{Geoga-etal-2022}
Geoga, C.J., Marin, O., Schanen, M. and Stein, M.L. (2022), Fitting Mat\'ern Smoothness Parameters
Using Automatic Differentiation, {\it arXiv:2201.00090v1}.

\bibitem{Gradshteyn2000}
Gradshteyn, T.S. and Ryzhik, I.M. (2000), {\it Table of Integrals, Series, and Products},
6th edition. Academic Press.

\item
Guttorp, P. and Gneiting, T. (2006), Studies in the History of Probability and Statistics XLIX.
On the Mat\'ern Correlation Family, {\it Biometrika}, 93, 989-995.

\bibitem{Handcock1993}
Handcock, M.S. and Stein, M.L. (1993), A Bayesian Analysis of Kriging, {\it Technometrics}, 35, 403-410.

\bibitem{Kaufman2013}
Kaufman, C.G. and Shaby, B.A. (2013), The Role of the Range Parameter for Estimation 
and Prediction in Geostatistics, {\it Biometrika}, 100, 473-484.

\bibitem{Keener2010}
Keener, R.W. (2010), {\it Theoretical Statistics: Topics for a Core Course}, Springer.

%\item
%Kim, T.Y., Park, J.S. and Song, G.M. (2010), An Asymptotic Theory for the Nugget Estimator in 
%Spatial Models, {\it Journal of Nonparametric Statistics}, 22, 181-195.

\bibitem{Loh2015}
Loh, W.-L. (2015), Estimating the Smoothness of a Gaussian Random Field From Irregularly Spaced 
Data Via Higher--Order Quadratic Variations, {\it Annals of Statistics}, 43, 2766-2794.


\bibitem{Loh-etal2021}
Loh, W.-L., Sun, S. and Wen, J. (2021), On Fixed--Domain Asymptotics, Parameter Estimation
and Isotropic Gaussian Random Fields With Mat\'ern Covariance Functions,
{\it The Annals of Statistics}, 49, 3127-3152.

%\item
%Loh, W.-L. (2005), Fixed--Domain Asymptotics for a Subclass of Mat\'ern--Type Gaussian 
%Random Fields, {\it Annals of Statistics}, 33, 2344-2394.

%\bibitem{Matern1986}
%Mat\'ern, B. (1986), {\it Spatial Variation}, 2nd. edition. Springer--Verlag.

\bibitem{McCulloch1989}
McCulloch, R.E. (1989), Local Model Influence,
{\it Journal of the American Statistical Association}, 84, 473-478.

\bibitem{Papritz2021}
Papritz, A. and Schwierz, C. (2021), {\tt georob}: Robust Geostatistical Analysis of Spatial Data,
{\tt R} package version 0.3-14, URL {\tt https://CRAN.R-project.org/package=georob}.

\bibitem{Seber2003}
Seber, G.A.F. and Wild, C.J. (2003), {\it Nonlinear Regression}, Wiley.

%\item
%Stein, M.L. (2010), Asymptotics for Spatial Processes, In: {\it Handbook of Spatial Statistics}, 
%A.E.. Gelfand, P.J. Diggle, M. Fuentes and P. Guttorp (eds.), CRC Press, pp 79-88.

\bibitem{Stein1999}
Stein, M.L. (1999), {\it Interpolation of Spatial Data: Some Theory for Kriging}. 
Springer--Verlag.

\bibitem{Stein1993}
Stein, M.L. (1993), A Simple Condition for Asymptotic Optimality of Linear Predictions of
Random Fields, {\it Statistics \& Probability Letters}, 17, 399-404.

\bibitem{Stein1990}
Stein, M.L. (1990), Uniform Asymptotic Optimality of Linear Predictions of a Random Field
Using an Incorrect Second--Order Structure, {\it Annals of Statistics}, 18, 850-872.

%\item
%Stein, M.L. (1990b), Bounds of Efficiency of Linear Predictions Using an Incorrect 
%Covariance Function, {\it Annals of Statistics}, 18, 1116-1138.

\bibitem{Stein1988}
Stein, M.L. (1988), Asymptotically Efficient Prediction of a Random Field With a Misspecified 
Covariance Function, {\it Annals of Statistics}, 16, 55-63.

\bibitem{Steiner1995}
Steiner, M., Houze, R.A. and Yuter, S.A. (1995), Climatological Characterization of Three--Dimensional
Structure From Operational Radar and Gauge Data, {\it Journal of Applied Meteorology}, 34, 1778-2007.

\item
Wu, W.-Y. and Lim (2016), Estimation of Smoothness of a Stationary Gaussian Random Field, 
{\it Statistica Sinica}, 26, 1729-1745.

\bibitem{Wu2013}
Wu, W.-Y., Lim, C.Y. and Xiao, Y. (2013), Tail Estimation of the Spectral Density for a 
Stationary Gaussian Random Field, {\it Journal of Multivariate Analysis}, 116, 74-91.

%\bibitem{Yaglom1987}
%Yaglom, A.M. (1987), {\it Correlation Theory of Stationary and Related Random Functions I:
%Basic Results}. Springer--Verlag.

\bibitem{Zhang2004}
Zhang, H. (2004), Inconsistent Estimation and Asymptotically Equal Interpolations
in Model--Based Geostatistics, 
{\it Journal of the American Statistical Association}, 99, 250-261.

\bibitem{Zhu-Zhang2006}
Zhu, Z. and Zhang, H. (2006), Spatial Sampling Design Under the Infill Asymptotic Framework, 
{\it Environmetrics}, 17, 323-337.

\bibitem{Zhu-Stein2006}
Zhu, Z. and Stein, M.L. (2006), Spatial Sampling Design for Prediction With Estimated Parameters, 
{\it Journal of Agricultural, Biological, and Environmental Statistics}, 11, 24-44.

\bibitem{Zhu-Stein2005}
Zhu, Z. and Stein, M.L. (2005), Spatial Sampling Design for Parameter Estimation of the 
Covariance Function, {\it Journal of Statistical Planning and Inference}, 134, 583-603.

\bibitem{Zimmerman2010}
Zimmerman, D.L. (2010), Likelihood--Based Methods, In: {\it Handbook of Spatial Statistics}, 
A.E. Gelfand, P.J. Diggle, M. Fuentes and P. Guttorp (eds.), CRC Press, pp 45-56.

\bibitem{Zimmerman2006}
Zimmerman, D.L. (2006), Optimal Network Design for Spatial Prediction, Covariance Parameter 
Estimation, and Empirical Prediction, {\it Environmetrics}, 17, 635-652.

\end{thebibliography}

\iffalse

\fi

\end{document}